\documentclass[sigconf]{acmart}
\AtBeginDocument{%
  \providecommand\BibTeX{{%
    \normalfont B\kern-0.5em{\scshape i\kern-0.25em b}\kern-0.8em\TeX}}}

\setcopyright{none}
\renewcommand\footnotetextcopyrightpermission[1]{}

\usepackage{algorithm}
\usepackage{graphicx}
\usepackage{amsmath}
\usepackage{amsthm}
\usepackage{algorithmic}
\usepackage{array}
\usepackage{subcaption}
\usepackage{gensymb}
\usepackage{url}
\usepackage{multicol}
\usepackage{multirow}
\usepackage{float}
\usepackage{framed}
\usepackage{xcolor}
\usepackage{mathtools}
\usepackage{etoolbox}

\newrobustcmd{\bb}{\bfseries}

\begin{document}

\title{\textit{Wiggle}: Physical Challenge-Response Verification of \\Vehicle Platooning}

 \author{Connor Dickey}
 \authornote{The authors are undergraduate students who equally contributed to this work during a 2021 summer REU at the University of Arizona.}
 \affiliation{%
   \institution{Bradley University}
   \country{}
 }
 
 \author{Christopher Smith\footnotemark[1]}
 \affiliation{%
  \institution{Stony Brook University}
  \country{}
}

 \author{Quentin Johnson\footnotemark[1], Jingcheng Li, Ziqi Xu, Loukas Lazos, Ming Li}
 \affiliation{%
  \institution{University of Arizona}
  \country{}
}

\renewcommand{\shortauthors}{}

\begin{abstract}
Autonomous vehicle platooning  promises many benefits such as fuel efficiency, road safety, reduced traffic congestion, and passenger comfort. Platooning vehicles travel in a single file, in close distance, and at the same velocity. The platoon formation is autonomously maintained by a Cooperative Adaptive Cruise Control (CACC) system which relies on sensory data and vehicle-to-vehicle (V2V) communications. In fact, V2V messages play a critical role in shortening the platooning distance while maintaining safety. Whereas V2V message integrity and source authentication can be verified via cryptographic methods, establishing the truthfulness of the message contents is a much harder task.

This work establishes a physical access control mechanism to restrict V2V messages to platooning members. Specifically, we aim at tying the digital identity of a candidate  requesting to join a platoon to its physical trajectory relative to the platoon. We propose the {\em Wiggle} protocol that employs a physical challenge-response exchange to prove that a candidate requesting to be admitted into a platoon actually follows it. The protocol name is inspired by the random longitudinal movements that the candidate is challenged to execute. {\em Wiggle} prevents any remote adversary from joining the platoon and injecting fake CACC messages. Compared to prior works, {\em Wiggle} is resistant to pre-recording attacks and can verify that the candidate is directly behind the verifier at the same lane. 
\end{abstract}


\maketitle
\pagestyle{plain}

\section{Introduction}

Autonomous platooning refers to the coordination of a group of autonomous vehicles traveling on a single file and in close proximity across long distances. Platooning offers notable benefits in road capacity and fuel efficiency due to the smaller inter-vehicle gaps, while maintaining safety \cite{alam2015heavy, lioris2017platoons}. The inter-vehicular gaps are maintained in a coordinated fashion without any mechanical linkage \cite{maiti2017conceptualization}.  Specifically, steering and acceleration is coordinated  using  vehicle-to-vehicle (V2V) communications and on-board sensors (cameras, LIDAR, radar). The platoon members employ a cooperative adaptive cruise control (CACC) algorithm to maintain safe distances within the platoon formation and react to the surrounding traffic \cite{turri2016cooperative,lyamin2016study, wang2018review}. V2V messages propagate much faster than sensory information that may lag in detecting imminent changes to vehicle trajectories. For instance, when one member brakes, this information propagates via a V2V message before a velocity change can be sensed. This allows for a considerable reduction of the safety gap between vehicles \cite{turri2016cooperative}.

\begin{figure}[t]
\centering
\includegraphics[width=0.9\columnwidth]{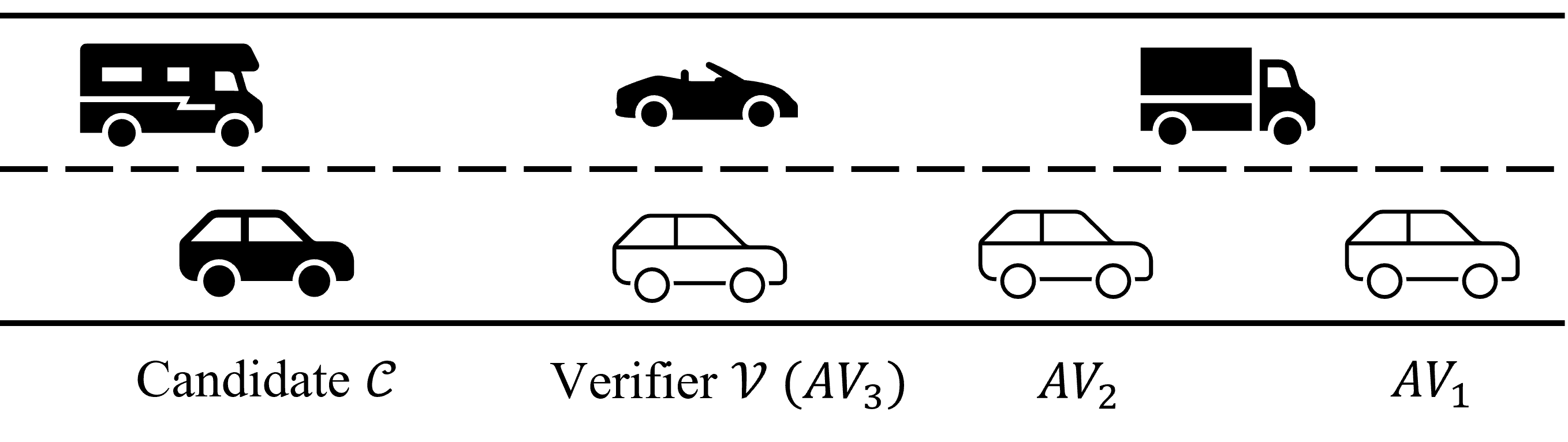}
\vspace{-0.1in}
\caption{A platoon of three vehicles formed by $AV_1$, $AV_2$, and $AV_3$. Vehicle $AV_3$ acts as a verifier $\mathcal{V}$ for the candidate $\mathcal{C}$ who wishes to be admitted to the platoon. Parties $\mathcal{C}$ and $\mathcal{V}$ engage in a Proof-of-Following protocol.}
\label{fig:systemmodel}
\vspace{-0.2in}
\end{figure}

The safety of the platoon and of other surrounding vehicles hinges on the veracity of V2V CACC messages. False message injection can lead to loss of life, monetary damages, and eventual abandonment of the autonomous platooning technology \cite{han2017convoy}. 
To secure the message exchange, wireless standards such as the IEEE 1609.2 \cite{IEEE:WAVE} and the more recent 3GPP TS 33.185 for Cellular Vehicle-to-Everything (C-V2X) \cite{secureV2X} recommend the use of a public key infrastructure (PKI). Cryptographic methods can authenticate the source and verify the integrity of a V2V message. However, they cannot physically bind the message originator to a trajectory. 

The lack of physical trajectory verification opens the door to remote attacks. An adversary could claim to follow a platoon while being at a remote location. Communication with the platoon may take place over the cellular infrastructure using C-V2X. The adversary may be in possession of valid cryptographic credentials either by being a valid vehicle or by compromising the credentials of valid vehicle. After authentication of the adversary's digital identity, the adversary can inject fake messages into the platoon and impact the CACC operation. This attack can scale to multiple platoons, as the adversary can simultaneously impersonate phantom vehicles at various distant locations.   

To mitigate the risks from remote attacks, several prior works have proposed physical access control mechanisms \cite{han2017convoy, xu2021pof, vaas2018get, juuti2017stash}. The main idea is to limit platoon access to only those vehicles that can prove they are actually following the platoon. The concept was formalized by Xu {\em et al.} with the introduction of a Proof-of-Following (PoF), which is demonstrated in Fig~\ref{fig:systemmodel}. Before being admitted to the platoon, a candidate member $\mathcal{C}$ engages in a challenge-response protocol with a verifier $\mathcal{V}$ (typically the last vehicle of the platoon) to bind $\mathcal{C}$'s digital identity with his physical trajectory. A PoF acts as a complementary mechanism to digital authentication by providing physical access control. As such, it does not prevent the injection of fake messages from vehicles that are already part of the platoon.  

{\bf Limitations of prior methods.}  The {\em Convoy} protocol uses the vertical acceleration due to road surface variations to correlate the  candidate's and the verifier's trajectories \cite{han2017convoy}. However, the road surface is static making {\em Convoy} vulnerable to pre-recording attacks. Our prior work in \cite{xu2021pof} exploits the large-scale fading effect of ambient cellular transmissions to correlate the candidate-verifier trajectories. This context presents high spatial and temporal entropy, thus resisting pre-recording attacks. However, the protocol cannot verify the relative positioning between the candidate and the verifier, but only bounds the candidate within a radius from the verifier. As a result the candidate could be anywhere within a radius.

In this paper, we propose {\em Wiggle,} a new PoF protocol that can verify (a) the following distance of the candidate, (b) the relative positioning of the candidate and the verifier, (c) the candidate's lane, and (d) provide resistance to pre-recording attacks. {\em Wiggle} employs a physical challenge-response exchange between the candidate and the verifier to prove that the candidate follows the platoon. The protocol name is inspired by the random longitudinal movements that the candidate is challenged to execute. Without following the verifier within the designated distance, a remote adversary cannot respond to the verifier's random motion challenges.

\textbf{Contributions:} Our main contributions are as follows:
\begin{itemize}
    \item We propose {\em Wiggle,} a PoF protocol that derives security form a series of physical challenges. These challenges are designed to bind the digital identity of the candidate to his trajectory, thus providing physical access control to the platoon. A physical challenge consists of a randomly-selected checkpoint that must be reached within a given deadline. Without following the platoon, a remote adversary cannot reach the checkpoints to prove his trajectory.
    
    \item We analyze the security of {\em Wiggle} and show that it is resistant to attacks from any malicious candidate that does not follow the verifier within a designated following distance $d_{ref},$ is not on the same lane as the verifier, or is separated by any other vehicle. Our protocol proves the relative ordering of the candidate and the verifier and provides lane verification. It is further resistant to pre-recording attacks due to the random nature of the physical challenges, and it is resistant to Man-in-the-middle (MitM) attacks when the identity of the verifier is known to the candidate. 
    
    \item We evaluate the performance and security of {\em Wiggle} via the Plexe platooning simulator \cite{segata2014plexe} and show that a PoF verification lasts less than a minute for relevant freeway scenarios, while providing a high security level. Moreover, by using an ACC algorithm to execute the challenges, we ensure that the user experiences almost imperceptible changes to the vehicle's velocity while a PoF is executed. 
\end{itemize}

\section{System Model}

\subsection{Platooning Model}
\label{sec:PlatooningModel}
We consider a vehicular platoon traveling on a single file in a freeway. The platoon members are either autonomous or semi-autonomous and coordinate driving via physical sensing and exchanging  V2V control messages that contain motion state information such as acceleration, velocity, steering, etc.  \cite{jia2015survey}. Vehicles are equipped with distance measuring sensors that are implemented using any modality such as radar, camera, LIDAR, or a combination of modalities \cite{yeong2021sensor}. Distance sensors are able to measure the distance to proceeding and following vehicles travelling in the same lane. Using the distance measurements and the exchanged messages, the platoon applies cooperative adaptive cruise control (CACC) \cite{turri2016cooperative,lyamin2016study} to maintain the platooning distance.

To secure the platoon operation, V2V messages are protected using cryptographic primitives. According to the  C-V2X  communication standard (3GPP TS 33.185 \cite{secureV2X}), V2X communication is supported by a PKI that provides each vehicle $X$ with a private/public key pair $(pk_X, sk_X)$ and a digital certificate $cert_X$. These credentials can be used to establish trust among the platoon vehicles and verify the origin of information. Key management of digital identities and platoon secrets is beyond the scope of this work.

\subsection{Platoon Physical Access Control} 

We study the problem of physical access control for securing vehicular platoons from remote adversaries. The main idea is to restrict platoon membership to those vehicles that are actually platooning, thus preventing remote adversaries from injecting fake navigation messages.   Figure \ref{fig:systemmodel} demonstrates our system model. Vehicles $AV_1$, $AV_2$ and $AV_3$ form a platoon. Candidate vehicle $\mathcal{C}$ requests to join the platoon claiming to be following $AV_3$ within the platooning distance. Vehicle $AV_3$ acts as a {\em verifier} for $\mathcal{C}$'s trajectory. 

\textbf{Candidate ($\mathcal{C}$):}
A candidate  $\mathcal{C}$ requests to be admitted to a platoon by sending a join request.  The candidate has a public/private key pair $(pk_\mathcal{C}, sk_\mathcal{C})$ and a certificate $cert_\mathcal{C}$ issued by a trusted certificate authority. The candidate is equipped with an adaptive cruise control (ACC) system that can autonomously maintain the following distance. The ACC can also become cooperative by receiving V2V messages. The candidate is not admitted to the platoon until it passes a PoF challenge from the verifier. 

\textbf{Verifier ($\mathcal{V}$):}
The last vehicle of a platoon serves as a verifier. The verifier engages in a PoF protocol with the candidate to  bind the candidate's digital identity with his physical trajectory. Like all other vehicles, the verifier is assigned a public/private key pair $(pk_\mathcal{V}, sk_\mathcal{V})$ and a certificate $cert_\mathcal{V}$ by a trusted certificate authority. Moreover, the verifier can securely measure the distance to any vehicle that follows within the same lane.  

{\bf The Proof-of-Following Primitive:}  A candidate should follows the platoon at the specified following distance $d_{ref}$ and be in the same lane as the verifier. We provide a stricter PoF definition than the one initially introduced in  \cite{xu2021pof}.

\begin{definition}
{\bf Route:} A route $\mathcal{L}_X$ of a vehicle $X$ is represented as a set of $n$ time-ordered positions $\mathcal{L}_X=(\ell_X(1), t(1)), (\ell_X(2),$ $(t(2)),\ldots,(\ell_X(n), t(n)),$ where each position $\ell_X(i)$ is the vehicle's geo-spatial coordinate $(x_X(i), y_X(i))$ at time $t(i),$ with $1\leq i \leq n$, and $t(i)<t(j)$ for $i<j$. 
\end{definition}

\begin{definition}
{\bf Proof-of-Following:} Let the verifier $\mathcal{V}$ move along a route $\mathcal{L}_V$ and a candidate $\mathcal{C}$ follow along a route $\mathcal{L}_C.$ If the candidate is following the verifier in the same lane and the Euclidean distance between $\mathcal{V}$ and $\mathcal{C}$ satisfies \vspace{-0.05in}
\[
||\ell_{\mathcal{V}}(i) - \ell_{\mathcal{C}}(i)|| = d_{ref},~~~\forall~i,
\] 
where $d_{ref}$ is a desired following distance, then $\mathcal{V}$ ACCEPTS. Else, the verifier REJECTS.
\label{PoF}
\end{definition}

\subsection{Threat Model}
\label{ssec:threat}

\textbf{Adversary goals and capabilities:}
We consider an adversary $\mathcal{M}$ who attempts to pass the PoF verification without following the platoon. The goal of the adversary is to be admitted into the platoon and inject false coordination messages. The attacker holds a public/private key pair $(pk_M, sk_M)$ and a certificate $cert_M$ issued by a trusted certificate authority. The adversary can communicate with the platoon either via C-V2X communications or directly. 
\begin{figure}[t]
\centering
\setlength{\tabcolsep}{-2pt}
\begin{tabular}{c}
  \includegraphics[width=0.9\columnwidth]{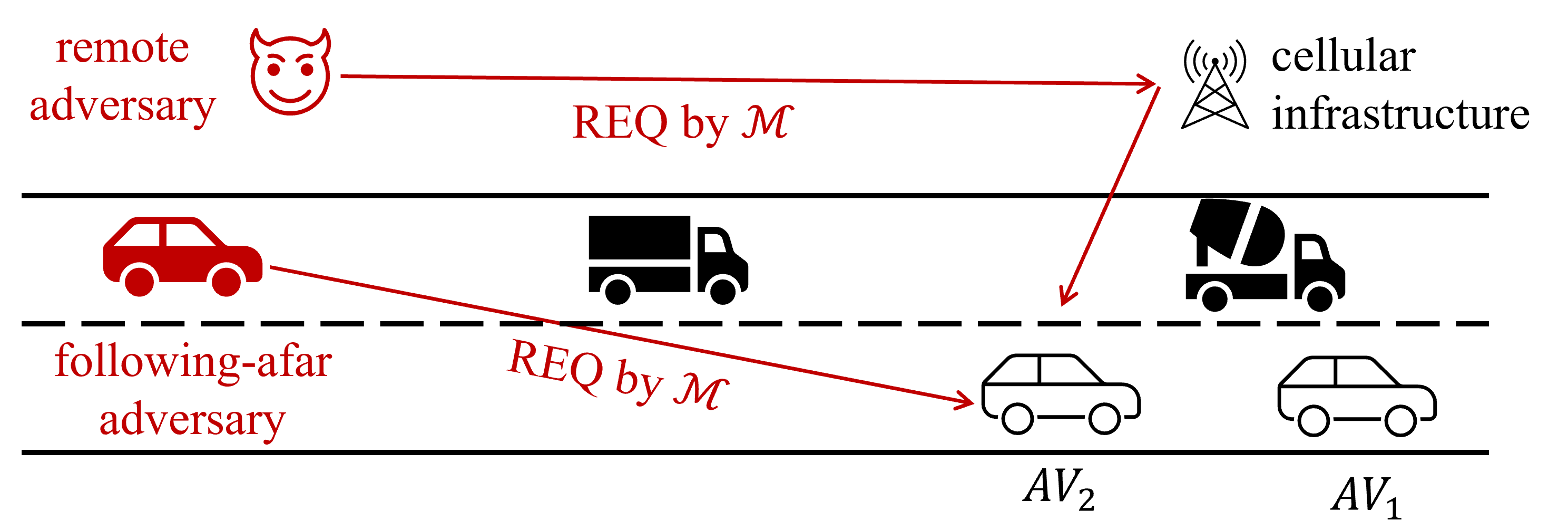}\\
  (a) a remote adversary\\
  \includegraphics[width=0.9\columnwidth]{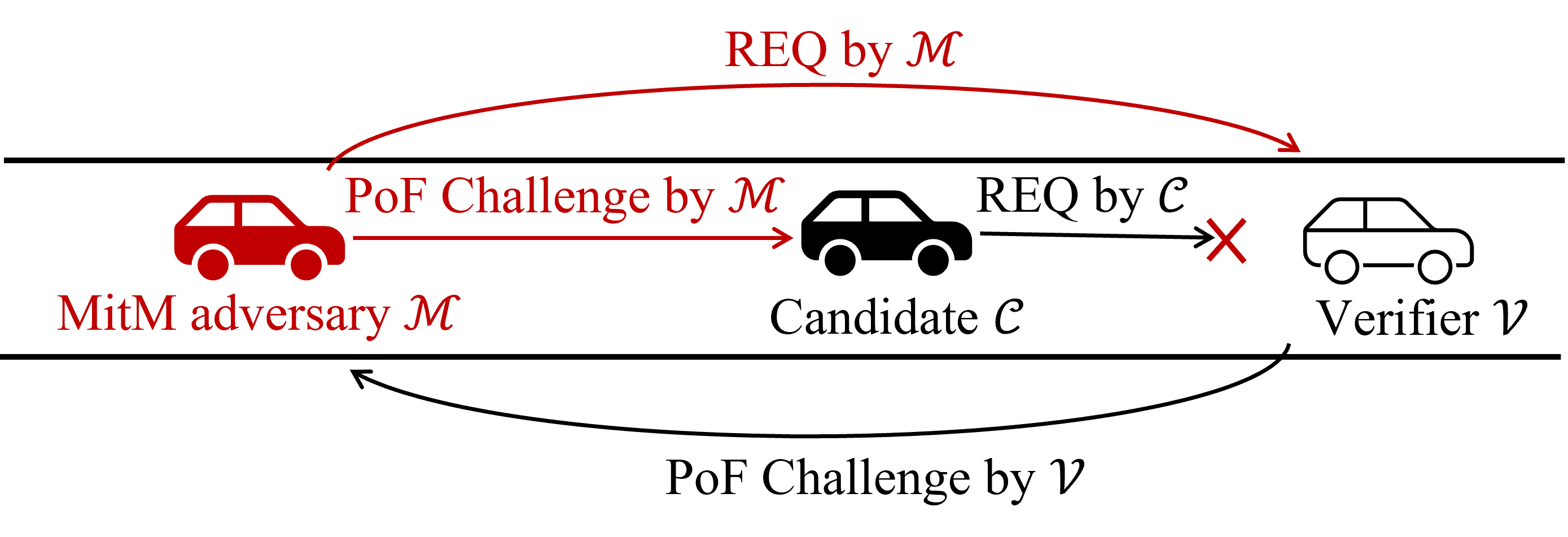}\\
  (b) a MiTM adversary  
\end{tabular}
\vspace{-0.1in}
\caption{The threat model.}
\label{fig:threatmodel}
\vspace{-0.2in}
\end{figure}

A remote adversary is shown in Fig. \ref{fig:threatmodel}(a). The adversary may know the platoon's route in advance or in real-time. He is using the cellular infrastructure to emulate a phantom vehicle which appears to follow the platoon. A similar scenario occurs when the adversary is at some far-away following distance or ahead of the platoon. Because the adversary is assumed to be remote, he does not launch attacks against the verifier's ranging sensors. Even if such attacks were launched, a secure ranging protocol can be used to protect the distance sensing modality \cite{singhVrange2022}.

{\em Man-in-the-middle adversary.} The adversary can launch a Man-in-the-Middle (MitM) attack to gain admittance to the platoon while a legitimate candidate $\mathcal{C}$ also attempts to join.
A MiTM attack is shown in Fig. \ref{fig:threatmodel}(b). The adversary jams the platoon join request sent from $\mathcal{C}$ and replaces it with his own request. At the same time, $\mathcal{M}$ impersonates the verifier to $\mathcal{C}$. The legitimate verifier challenges $\mathcal{M}$ to prove it follows the platoon by executing a PoF. The adversary relays the same challenge to $\mathcal{C}$ who executes the PoF protocol.

\section{The \textit{Wiggle} PoF Protocol}

\subsection{Overview}
\textit{Wiggle} is a physical challenge-response protocol executed between the verifier and candidate. To bind the digital identity of a candidate with his physical trajectory, the verifier challenges the candidate to execute a series of longitudinal perturbations of its following distance and measures these perturbations using the ranging modality. The physical challenges are randomly generated by the verifier and sent to the candidate  encrypted by the candidate's public key. Each challenge $(d_i, t_i)$ consists of a desired following distance $d_i,$ referred to as a ``checkpoint'', and a deadline $t_i$.

\begin{figure}[t]
\centering
\includegraphics[width=0.9\columnwidth]{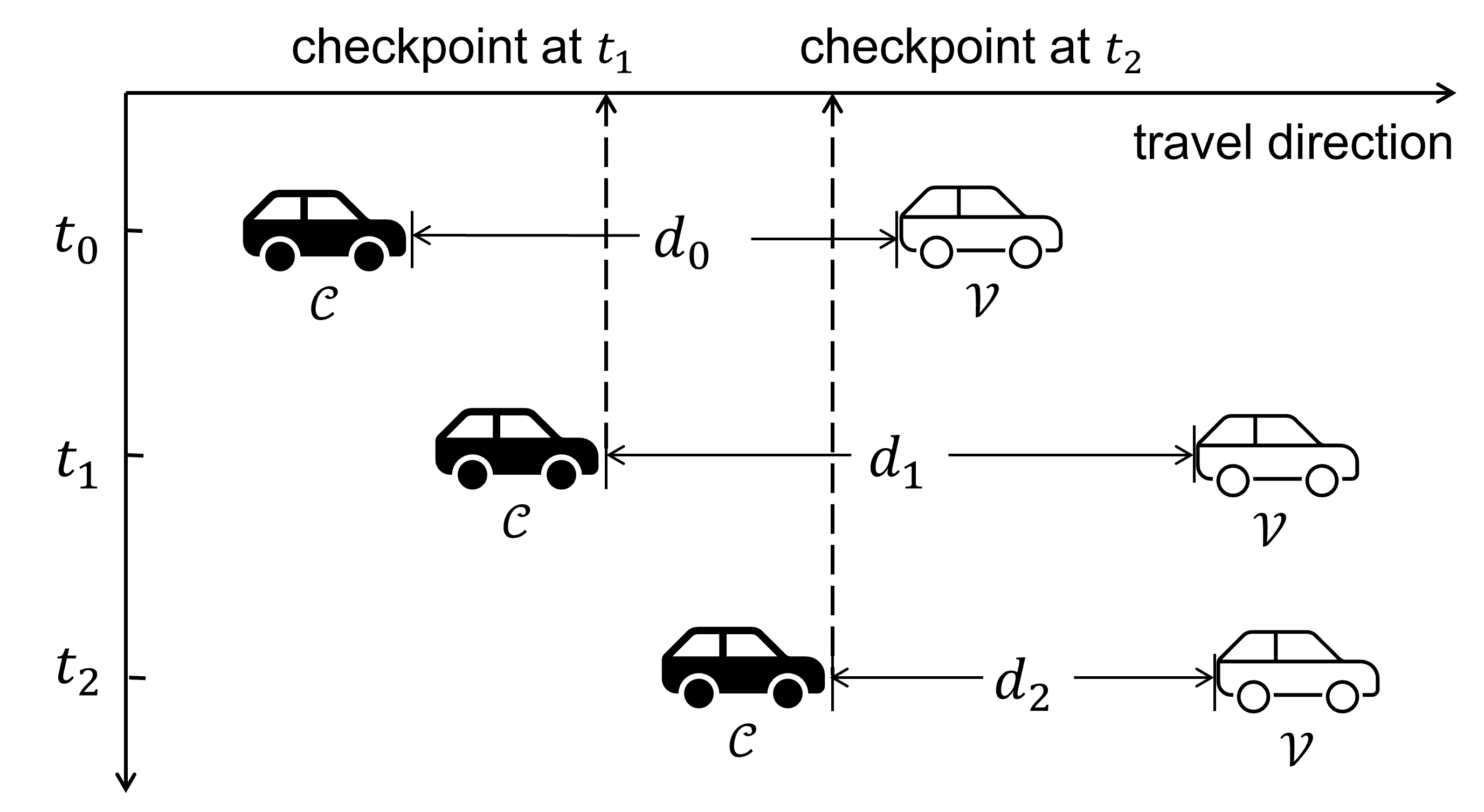}
\vspace{-0.1in}
\caption{The verifier challenges the candidate to reach randomly-generated checkpoints $d_1$ and $d_2$ by deadlines $t_1$ and $t_2,$ respectively.}
\label{fig:checkpoint}
\vspace{-0.2in}
\end{figure}

Figure \ref{fig:checkpoint} shows the execution of two challenges by candidate $\mathcal{C}$. At time $t_0$, $\mathcal{C}$ claims to follow $\mathcal{V}$ at following distance $d_0 = d_{ref}.$ The verifier measures the distance to the following vehicle and verifies that it is $d_{ref}$. However, this alone does not constitute a PoF as another vehicle could happen to follow $\mathcal{V}.$ The verifier challenges $\mathcal{C}$ to reach checkpoints $d_1$ and $d_2$ by deadlines $t_1$ and $t_2$, respectively. The challenges are encrypted by $\mathcal{C}$'s public key. To pass PoF verification, the candidate must reach each checkpoint by the designated deadline, resulting in a ``wiggle'' motion around the following distance $d_{ref}.$ A remote adversary is unable to pass verification, as he cannot be present at the checkpoints by the designated deadlines. Furthermore, by pointing the ranging sensor directly behind the verifier, relative ordering verification and lane verification are achieved.

\begin{figure*}[t]
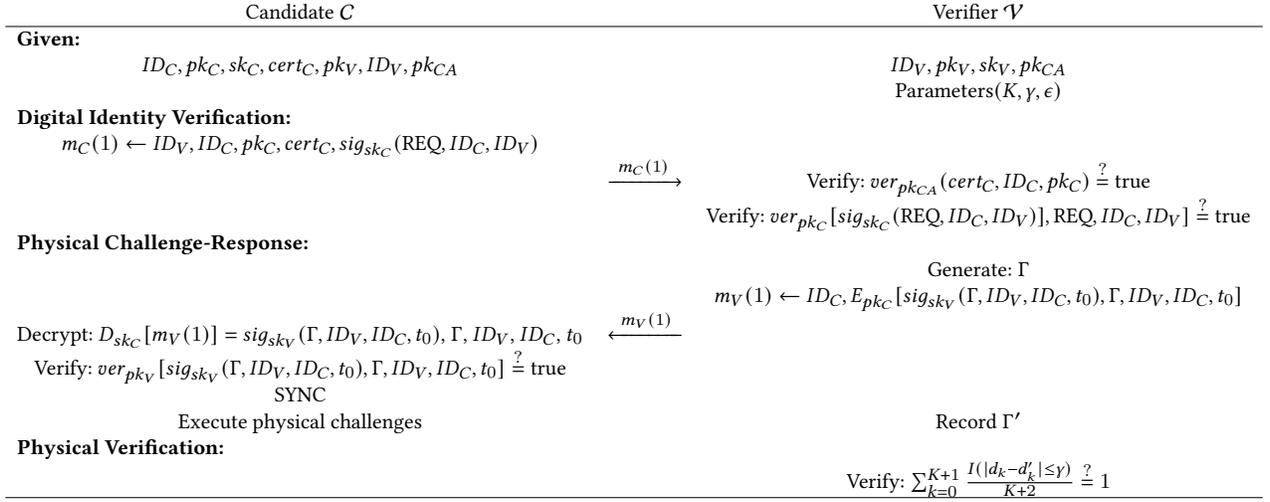

\centering
\scalebox{.9}{
    \begin{tabular}{ccc}
    Candidate $\mathcal{C}$ & & Verifier $\mathcal{V}$  \\
    \hline
    \multicolumn{1}{l}{\bf Given:} & &\\
        $ID_{C},pk_{C},sk_{C},cert_C, pk_{V}, ID_V, pk_{CA}$ & & $ID_V, pk_V, sk_V, pk_{CA}$ \\
        & & Parameters$(K, \gamma, \epsilon)$ \\
             
    \multicolumn{1}{l}{{\bf Digital Identity Verification:}}& &\\
        $m_C(1) \leftarrow ID_V,ID_C, pk_C, cert_C, sig_{sk_C}(\text{REQ}, ID_C, ID_V)$ & & \\
        
        &$\xrightarrow{~m_{C}(1)~}$ & Verify: $ver_{pk_{CA}}(cert_C,ID_C,pk_C)\overset{?}{=} \text{true}$\\
        
        & & Verify: $   ver_{pk_C}[sig_{sk_C}(\text{REQ}, ID_C, ID_V)],\text{REQ}, ID_C, ID_V] \overset{?}{=} \text{true}$  \\
      \multicolumn{1}{l}{{\bf Physical Challenge-Response:}}& &\\
      
        & & Generate: $\Gamma$\\
        
        & & $m_V(1) \leftarrow ID_C, E_{pk_C}[sig_{sk_V}(\Gamma, ID_V, ID_C, t_0), \Gamma, ID_V, ID_C, t_0]$\\
        
        Decrypt: $D_{sk_C}[m_V(1)] = sig_{sk_V}(\Gamma, ID_V, ID_C, t_0)$, $\Gamma$, $ID_V$, $ID_C$, $t_0$ &  $\xleftarrow{~m_{V}(1)~}$& \\
            
        Verify: $ver_{pk_V}[sig_{sk_V}(\Gamma, ID_V, ID_C, t_0),\Gamma, ID_V, ID_C, t_0] \overset{?}{=} \text{true}$ & &\\     
        SYNC & & \\
             
        Execute physical challenges  & & Record $\Gamma'$ \\
             
    \multicolumn{1}{l}{{\bf Physical Verification:}}& &\\
             
        & & Verify: $\sum_{k=0}^{K+1} \frac{I(|d_k -d'_k|\leq \gamma )}{K+2}\overset{?}{=} 1$ \\
    \hline
    \end{tabular} }
\caption{The \textit{Wiggle} protocol.}
\label{fig:protocol}
\vspace{-0.1in}
\end{figure*}

\subsection{The \textit{Wiggle} Protocol}
We organize the protocol into three phases: \textit{Digital identity verification}, \textit{Physical challenge-response}, and \textit{Physical Verification}. In the first phase, the digital identity of the candidate is verified. In the challenge-response phase, the verifier challenges the candidate to perform random motions and measures the distance to the candidate. In the physical verification phase, the verifier accepts the candidate's claim if he reaches all checkpoints by the respective deadlines.  Figure \ref{fig:protocol} summarizes the protocol in steps. 
\medskip

\noindent \textbf{\textit{Digital identity verification phase.}}  
\begin{enumerate}
    
  \item The candidate sends a join request REQ to the verifier $\mathcal{V}$. 
  \[
  m_C(1) \leftarrow ID_V,ID_C, pk_C, cert_C, sig_{sk_C}(\text{REQ}, ID_C, ID_V),
  \]
  where $ID_V,ID_C$ are the verifier's and the candidate's identities, $(pk_C, sk_C)$ are $\mathcal{C}$'s
public/private key pair, and $cert_C$ is $\mathcal{C}$'s certificate.  

  \item The verifier validates the certificate of $\mathcal{C}$ with  $pk_{CA}$  the certificate authority's public key $pk_{CA}$   and then verifies the signature with $pk_C$. 
  
  
\end{enumerate}

\textbf{\textit{Physical challenge-response phase.}}
\begin{enumerate}
  \setcounter{enumi}{3}
  
  \item $\mathcal{V}$ generates a set of $K$ physical challenges denoted by $\Gamma=\{(d_{ref},t_0), (d_1,t_1),\cdots,(d_K,t_K), (d_{ref},t_{K+1})\}.$ Each challenge  consists of a checkpoint $d_i$, which is a random longitudinal  perturbation of the following distance $d_{ref}$, and a corresponding deadline $t_i$ by which the checkpoint must be reached. The set also contains the initial and final position of $\mathcal{C}$ that is equal to $d_{ref}.$ The challenges are signed with $sk_V$ and then encrypted with $pk_C.$ The message also contains the start time $t_0$ of initiating the response. 
  \[
  m_V(1) \leftarrow ID_C, E_{pk_C}[sig_{sk_V}(\Gamma, ID_V, ID_C, t_0), \Gamma, ID_V, ID_C, t_0].
  \]
    
  \item $\mathcal{C}$ decrypts $m_V(1)$ and verifies the signature of $\mathcal{V}.$ 
  
  \item $\mathcal{C}$ starts from $d_{ref}$ at time $t_0$, passes through each checkpoint $d_i$ by deadline $t_i$ and then recovers to $d_{ref}$.
  
  \item The verifier measures and records the following distance of the candidate by each deadline. Denote the recorded data set as $\Gamma'=\{(d'_0, t_0), (d'_1, t_1),(d'_2,t_2),\cdots,(d'_K,t_K),(d'_{K+1}, t_{K+1}) \}.$ 
\end{enumerate}

\textbf{\textit{Physical verification phase.}}
In this phase, $\mathcal{V}$ verifies the candidates platooning \ claim by checking if $\mathcal{C}$ reached the designated checkpoints by the respective deadlines. 
\begin{enumerate}
  \setcounter{enumi}{5}
  
  \item $\mathcal{V}$ compares each  measured distance $d`_{i}$ with the respective challenge $d_i$. If each $d`_{i}$ is within a threshold $\gamma$ from  $d_i$, the verifier ACCEPTS. Otherwise, the verifier REJECTS.  
  \[
  \sum_{k=0}^{K+1} \frac{I(|d_k -d'_k|\leq \gamma)}{K+2} = 1,
  \]
  where $I(\cdot)$ is the indicator function.
\end{enumerate}

\subsection{Parameter Selection}
\label{sec:ACCcontrol}

{\bf Checkpoint selection.} To select each checkpoint $d_i$, the verifier determines a discrete range $\mathcal{S}$ around the nominal following distance $d_{ref}$. Using the standard time gap notation to denote following distances,  let $d_{ref}$ correspond to a time gap $g_{ref} = d_{ref}/v_V$, where $v_V$ denotes the verifier's velocity. Let also $g_{\min}$ to be the minimum safety time gap between any two vehicles and $g_{\max}$ be a maximum time gap. The verifier computes a continuous range $[g_{\min}\cdot v_V, g_{\max} \cdot v_V]$ for selecting the checkpoints. It then divides this range to equal segments of length $2\rho$ (twice the radar resolution $\rho$) and computes a discrete range of $M$ checkpoints $\mathcal{S}=\{s_1, s_2,\ldots,s_M\}$ where 
\[
M = \lfloor \frac{(g_{\max} - g_{min}) \cdot v_V}{2\rho}\rfloor + 1.
\]
The checkpoint for each challenge is randomly selected from $\mathcal{S}.$ 

\begin{figure}[t]
\centering
\setlength{\tabcolsep}{-3pt}
\begin{tabular}{c}
  \includegraphics[width=\columnwidth]{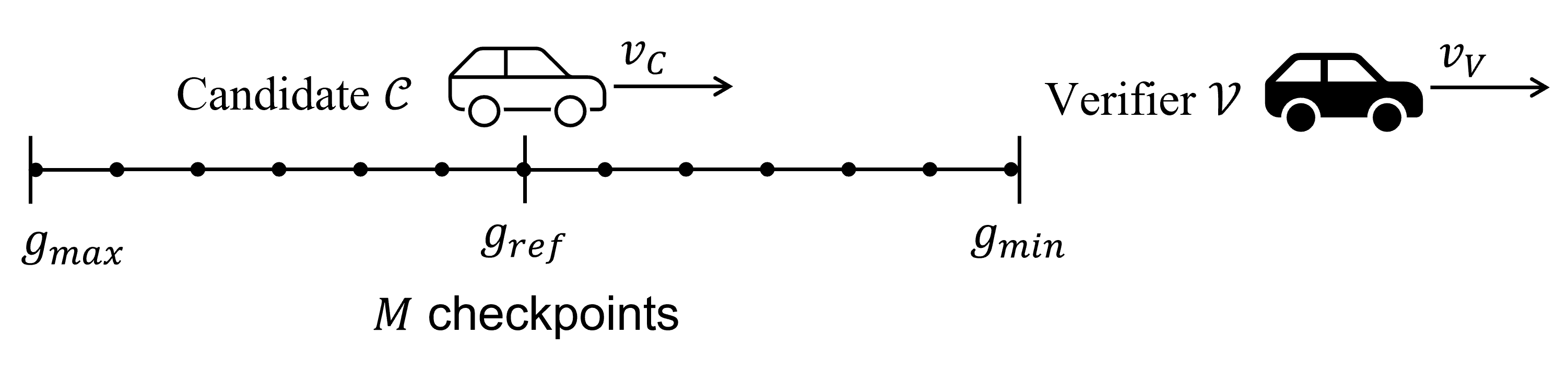}
\vspace{-0.2in}
\end{tabular}
\caption{Setting the checkpoint range for candidate $\mathcal{C}$.}
\label{fig:mstate}
\vspace{-0.2in}
\end{figure}

To demonstrate the checkpoint selection process, consider a verifier  traveling at $v_V= 30$m/s, as shown in Fig.~\ref{fig:mstate}.
Assume $g_{min}=1s$, $g_{max}=2s$ and a radar resolution of $\rho=0.3m$. The verifier computes  $M=\frac{60\text{m} - 30\text{m}}{2\cdot 0.3\text{m}}+1 =51$ checkpoints between 30m and 60m from itself.  The verifier randomly chooses from the 51 checkpoints when populating the $K$ physical challenges for any candidate.

{\bf Deadline selection.} The deadlines can be selected in any fashion that would allow the candidate to safely move to the designated checkpoints. A straightforward way to select a deadline $t_i$ for checkpoint $d_i$ is to assume some relative velocity differential $v_{rel}$ (positive or negative) to cover the distance difference $||d_i - d_{ref}||.$ In this case the deadline becomes $t_i = \frac{||d_i - d_{ref}||}{v_{rel}} + \epsilon$, where $\epsilon$ is some tolerance to allow for small variations in the candidate's motion. 

However, this simple model ignores the automated nature of platooning and the user experience, as the candidate's velocity is assumed to change instantly rather than smoothly. Alternatively, the verifier can calculate deadlines using an ACC model that accounts for safety and motion smoothness factors. Here, we adopt the ACC control model presented in  \cite{rajamani2011vehicle}, but any ACC controller can be used. Using this model, the deadline is calculated as follows. Let a challenge $d$ correspond to a gap time of $T=d/\dot{x}_C$ where $\dot{x}_C$ denotes the current speed of the candidate. The algorithm proceeds in steps of duration $\Delta_t$ as follows:

\begin{enumerate}

\item The desired acceleration at the $n$-th step is \vspace{-0.07in}
\begin{eqnarray}
  \ddot{x}_{des}[n] = -\frac{1}{T}(\Delta \dot{x}[n] + \lambda \delta[n]) \label{eq:acc_desire} \label{eq:acc_delta} \\
  \delta[n] = -d_{act}[n]+d, ~~~\Delta \dot{x}[n] = \dot{x}_C[n] - \dot{x}_V[n],
\end{eqnarray}
where $\Delta \dot{x}[n]$ is the relative velocity between $\mathcal{C}$ and $\mathcal{V}$, $d_{act}[n]$ is the actual following distance, $\delta[n]$ is the distance error to the desired checkpoint $d$, and $\lambda>0$ is a design parameter that controls the rate of convergence to $d$.

\item Instead of applying $\ddot{x}_{des}[n]$, the acceleration applied involves the input from the previous step:
\begin{align}
  \ddot{x}[n]  = \beta \cdot \ddot{x}_{des}[n] + (1-\beta) \cdot \ddot{x}[n-1],~~~~~
  \beta = \frac{\Delta_t}{\tau+\Delta_t}.
\end{align}
Here, $\tau$ is a time constant typically set to 0.5s and  $\Delta_t$ denotes the time gap between the $(n-1)$-st and $n$-th steps.
    
\item The distance gain of $\mathcal{C}$ during $\Delta_t$ is computed by
\begin{align}
  l[n] = \dot{x}[n-1] \cdot \Delta_t + \frac{1}{2} \cdot \ddot{x}[n] \cdot {\Delta_t}^2.
\end{align}    

\item The distance $\delta[n]$ to the checkpoint $d$ at step $n$ is updated to
\begin{align}
  \delta[n] = \delta[n-1]+l[n]-\dot{x}_{V}[n] \cdot \Delta_t.
\end{align}

\item Steps 1-4 are iterated through until  $|\delta[n]|<\gamma$ where $\gamma$ is the checkpoint distance tolerance. The deadline $t$ for a checkpoint $d$ is set to $t=\Delta t *n^{\ast}$, where $n^{\ast}$ is the first value of $n$ for which $|\delta[n]|<\gamma.$
    
\end{enumerate}

\section{Security Analysis}
\label{sec:sec_ana}

\subsection{Correctness} 
A valid candidate following the verifier at $d_{ref}$ will receive from $\mathcal{V}$ the set of challenges $\Gamma.$ Feeding $\Gamma$ to the ACC will enable $\mathcal{C}$ to reach the checkpoints by the deadlines while $\mathcal{V}$ measures his position, thus completing the PoF. Moreover, because $\mathcal{V}'s$ radar modality measures directly backwards within the same lane, both the relative ordering and the travelling lane are verified. 

\subsection{Remote Adversary}
\label{ssec:remote}

We first examine if an adversary that is not platooning is able to pass the PoF and join the platoon. The adversary could be at any location except $d_{ref}$ behind the verifier (for all practical purposes, an adversary following the platoon at $d_{ref}$ should be allowed to pass physical access control). For instance, the adversary could be stationary at a remote location, several cars behind the verifier, co-traveling at a different lane, etc. We consider two possibilities: (a) there are no other vehicles following the verifier and (b) a vehicle other than the adversary follows the verifier. 

\subsubsection{No vehicles follow $\mathcal{V}$}
Let the adversary $\mathcal{M},$ request to join the platoon by sending a request message $m_M(1)$ to the verifier. The adversary will pass the digital identity verification as he is assumed to posses a valid certificate issued by a trusted certificate authority.  After identity verification, the verifier will challenge the adversary $\mathcal{M}$ with a set of challenges $\Gamma.$ As the adversary does not follow $\mathcal{V}$ and no other vehicles follow $\mathcal{V}$, the verifier will be unable to detect a vehicle at the designated checkpoints, and the physical verification will fail.

\subsubsection{A vehicle follows $\mathcal{V}$}

We now consider the case where some vehicle $\mathcal{R}$ other than the adversary follows the verifier. The vehicle $\mathcal{R}$ is not controlled by the adversary, but is in the same lane as the verifier and keeps a safe distance that could be similar to the following distance $d_{ref}$. The remote adversary requests to join the platoon by sending a request message $m_M(1)$ to the verifier. As mentioned before, the adversary will pass the digital identity verification. The verifier will challenge the adversary with a set of challenges $\Gamma.$ The verifier will measure the distance to the following vehicle $\mathcal{R}$ (instead of the remote adversary) at the deadlines designated in $\Gamma.$  The adversary could pass the PoF, if $\mathcal{R}$ happens to be at the checkpoints by the respective deadlines.



\begin{figure}[t]
\centering
\setlength{\tabcolsep}{-3pt}
\begin{tabular}{c}
  \includegraphics[width=\columnwidth]{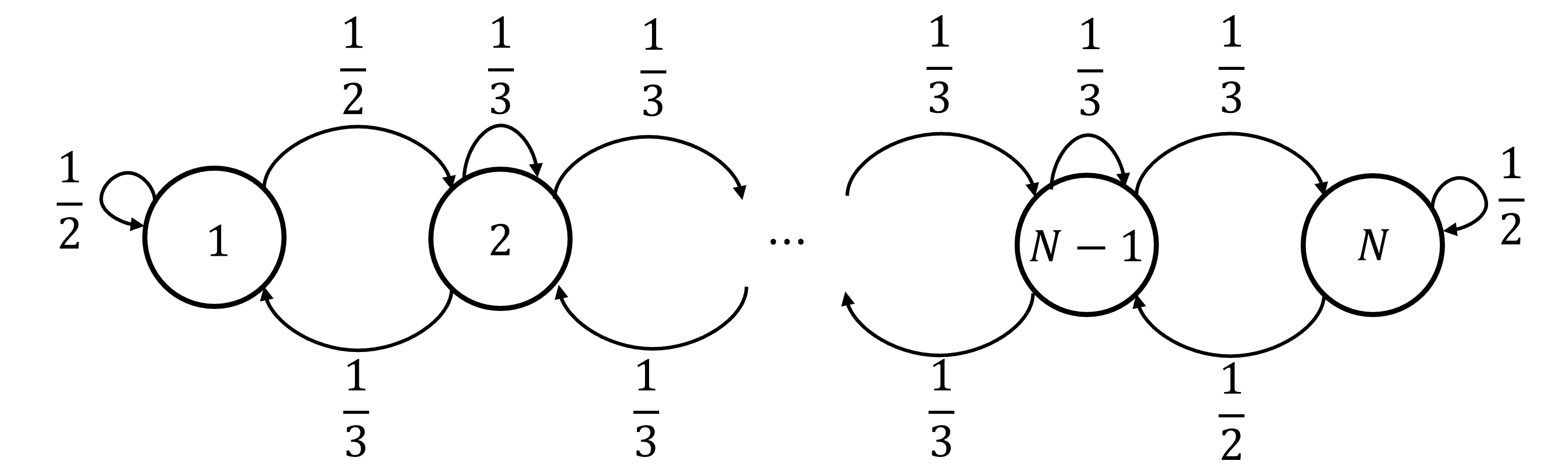}\\\vspace{-0.1in}
  (a) \\
  \includegraphics[width=\columnwidth]{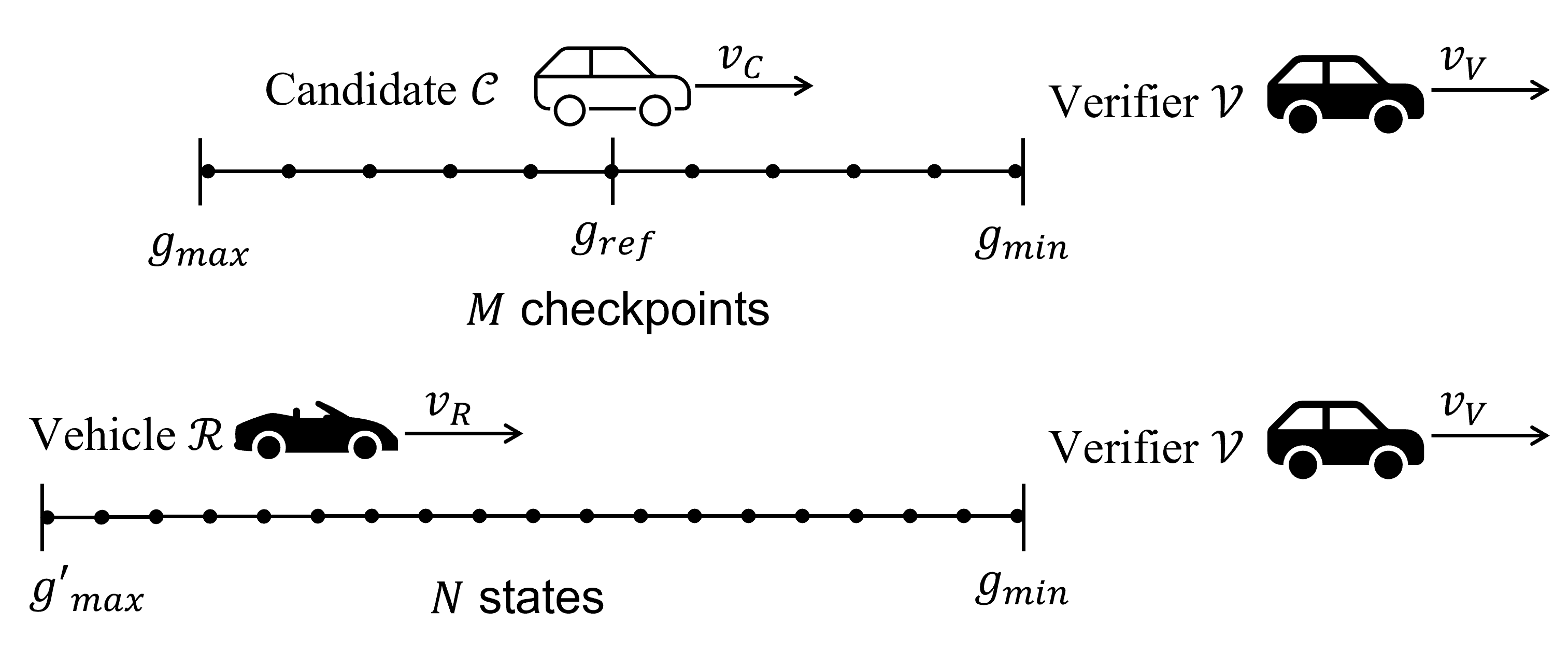}\\
  (b)
\vspace{-0.1in}
\end{tabular}
\caption{(a) The Markov chain model for the random walk of vehicle $\mathcal{R}$, (b) the $M$ checkpoints selected by the verifier and the $N$ possible states of vehicle $\mathcal{R}$.}
\label{fig:markovchain}
\vspace{-0.2in}
\end{figure} 

\textbf{Modeling $\mathcal{R}$'s trajectory as a random walk:} To analyze the probability of passing the PoF, we model the trajectory of the following vehicle $\mathcal{R}$ as a one-dimensional random walk around $d_{ref}.$ The core idea is that $\mathcal{R}$ moves independently of the platoon and may fluctuate its following distance within a limited range while still following. Specifically, the vehicle $\mathcal{R}$ fluctuates its distance to the verifier within a range $[d_{\min}, d_{\max}]$.

We model the random walk of $\mathcal{R}$ 
by an $N$-state Markov chain where states represent the candidate positions of $\mathcal{R}$ and state transition probabilities represent the probability of moving to  another position within the range after a time step $n$. We discretise the range $[d_{\min}, d_{\max}]$, by assuming that $\mathcal{R}$ can travel a fixed distance $d_{step}$ within a fixed time step and divide the range by $d_{step}$ to obtain a total of $N$ positions (states). Without loss of generality, the initial state distribution $P^{(0)}$ at time 0 is assumed to be uniform. Moreover the state transition probabilities are given by an $N\times N$ matrix $P =(P_{ij})$ with 
\begin{eqnarray}
P_{1,1} = P_{1,2} = P_{N,N}= P_{N,N-1}= 1/2, \\
P_{i,i+1} = P_{i,i-1} = P_{i,i} = 1/3,~~i = 2..N-1,\\
P_{i,j} = 0,~~\text{all other}~~i, j. \label{transitionProb}
\end{eqnarray}
The transition state diagram of the random walk is shown in Fig.~\ref{fig:markovchain}(a). Note that in a typical random walk, there is always a transition to a new state. In our model, we have opted to consider that the vehicle may stay on the same state within a time step. Moreover, given a state, the transition probabilities forward, backward, and at the same state are equiprobable, though any matrix $P$ can be considered. The $N$ candidate states of $\mathcal{R}$ may not necessarily coincide with the $M$ possible checkpoints selected by the verifier. However, we can assume that the $M$ checkpoints are part of the state space of $\mathcal{R}.$
Using the random walk model, we now evaluate the probability of passing the PoF verification, considering only the physical challenges and ignoring the initial and final  states of $d_{ref}$. 

\medskip
\noindent \textbf{Proposition 1.} \textit{Let the verifier challenge the  adversary $\mathcal{M}$  with a set of $K$ challenges $\Gamma = \{(d_1, t_1), (d_2, t_2),\ldots (d_K, t_K)\}.$ Each checkpoint is randomly selected from a state space $\mathcal{S}$ of size $M$. Let some vehicle $\mathcal{R}$ follow the verifier and move in a state space $\mathcal{S}'$ of size $N$ using the random walk model with $\mathcal{S} \subseteq \mathcal{S}'$.  The probability that $\mathcal{M}$ passes the PoF verification due to $\mathcal{R}$'s motion is given by}
\begin{align}
  \mathbf{P}_M =  \left(\frac{1}{NM}\right)^K \prod_{k=1}^K \sum_{i=1}^{M} \sum_{j=1}^N P^{\sum_{\ell=1}^k n_k}_{j,i}.
\label{eq:passingrate}
\end{align}
where $P^{n}$ indicates the transition probability matrix after $n$ steps.

\begin{proof}

{\bf Passing a single challenge:} Consider a single challenge with a  checkpoint/deadline pair $(d, t)$. The adversary passes the  challenge if $\mathcal{R}$ happens to be at distance $d$ during time $t$. The deadline $t$  corresponds to the $n$-th time step in the discrete stochastic process representing the random walk of $\mathcal{R}$ and captured by random variables $(X_1, X_2,\ldots)$ where 
\begin{align}
    X_{n}=X_0 \cdot P_0^n.
\end{align}

\noindent Denote by $\Pr[pass]$ the probability of reaching $d$ in time step $n$, or equivalently that the adversary passes the challenge. Then 
\begin{subequations}
\begin{align}
    \Pr[pass]
    & =\sum_{i = 1}^{M}\Pr[d = i] \cdot \Pr[X_{n}= i~|~ d= i] \label{pass1} \\
    & =\frac{1}{M} \sum_{i=1}^{M} \cdot \Pr[X_{n} =i~|~d=i]  \label{pass2}\\
    & =\frac{1}{M} \sum_{i=1}^{M} \sum_{j=1}^N Pr[X_0 = j] \cdot \Pr[X_{n}=i~|~X_0=j]  \label{pass3}\\
    &= \frac{1}{NM} \sum_{i=1}^{M} \sum_{j=1}^N P^{n}_{j,i}  
    \label{pass4}
\end{align}
\end{subequations}
In \eqref{pass1}, we conditioned on all the $M$ possible checkpoint values. In \eqref{pass2}, we considered that checkpoints for each challenge are chosen uniformly. In \eqref{pass3}, we conditioned on all the initial $N$ states for vehicle $\mathcal{R}$, and in \eqref{pass4} we considered that the initial location of vehicle $\mathcal{R}$ is uniform on the $N$ states in the random walk. 

{\bf Passing a PoF:} To pass a PoF challenge, the adversary must pass all $K$ challenges issued in set $\Gamma = \{(d_1, t_1), (d_2, t_2),\ldots (d_K, t_K)\}.$ In other words, $\mathcal{R}$ must reach each checkpoint by the designated deadline by means of a random walk. Since the verifier selects each checkpoint $d_k$ at random and independently, the passing rate can be expressed as
\begin{subequations}
\begin{align}
\mathbf{P}_M &= \Pr[\text{pass chal 1}] \ldots \Pr[\text{pass chal}~K] \label{total1} \\
 &= \left(\frac{1}{NM} \sum_{i=1}^{M} \sum_{j=1}^N P^{n_1}_{j,i} \right) \ldots \left(\frac{1}{NM} \sum_{i=1}^{M} \sum_{j=1}^N P^{n_1+n_2+\ldots+n_k}_{j,i}  \right) \label{total2} \\
 &= \left(\frac{1}{NM}\right)^K \prod_{k=1}^K \sum_{i=1}^{M} \sum_{j=1}^N P^{\sum_{\ell=1}^k n_k}_{j,i}. \label{total3}
\end{align}
\end{subequations}
Eq. \eqref{total1} follows by the independence of the challenges. In Eq. \eqref{total2}, we used the probability of passing a single challenge obtained by \eqref{pass4}, but accumulating the time steps from one challenge to another. Eq. \eqref{total3} is a more compact expression of \eqref{total2}. 
\end{proof}


\noindent \textbf{Lemma 1.} \textit{The adversary's passing probability  is upper bounded by}
\begin{align}
    P_M \leq \left(\frac{1}{M}\right)^K.
\label{eq:lemma}
\end{align}

\begin{proof}
To prove Lemma 1, we focus on the probability of passing a single challenge as expressed in \eqref{pass4}. We note that the summation term  $\sum_{i=1}^MP^{n}_{j,i}$ sums $M$ out of $N$ elements of the $j$-th row of matrix $P^{n}.$ Because the space of checkpoints is a subspace of the space of candidate locations (states) for vehicle $\mathcal{R},$ it follows that 
\[
\sum_{i=1}^M P^{n}_{j,i} \leq \sum_{i=1}^N P^{n}_{j,i} =1.
\]
Substituting this bound to Eq. \eqref{pass4} yields
\begin{align}
\Pr[pass] &= \frac{1}{NM} \sum_{i=1}^{M} \sum_{j=1}^N P^{n}_{j,i} 
\leq \frac{1}{NM}  \sum_{j=1}^N 1
= \frac{1}{M}. \notag
\end{align}
Substituting this bound to Eq. \eqref{total3} completes the proof. 
\end{proof}

From Lemma 1, we observe that the passing probability $P_M$ drops {\em at least} inversely proportional to the cardinality $M$ of the checkpoint space and exponentially with the number of challenges $K$. By controlling these two parameters, the passing probability can be driven to any desired value at the expense of delay until the PoF verification is completed.

We note that the bound of Lemma 1 is quite loose when $M<<N.$ In fact, one can show that as time accumulates, the steady-state distribution for the random walk becomes uniform (with the exception of the two boundaries that have different transition probabilities). Under a uniform distribution on any of the $N$ states, the probability of passing a single challenge becomes $\Pr[pass] =1\slash N$, which is independent of $M.$ That is, the probability of being at the selected checkpoint is one out of the possible $N$ states of the Markov chain. Given the $K$ independent challenges, the probability of passing PoF verification becomes $P_M \sim (1\slash N)^K.$

\subsection{A MiTM Adversary} In a MiTM attack, the adversary attempts to be admitted to the platoon when a valid candidate initiates a join request with the verifier. We analyze two instances of the attack. In the first instance, the candidate attempts to join a specific platoon with a known verifier identified by his public key $pk_V$ and his certificate $cert_V$. In the second case, the candidate opportunistically attempts to join a nearby platoon, without targeting a specific verifier. 

\begin{figure}[t]
\centering
\setlength{\tabcolsep}{-3pt}
\begin{tabular}{c}
  \includegraphics[width=0.9\columnwidth]{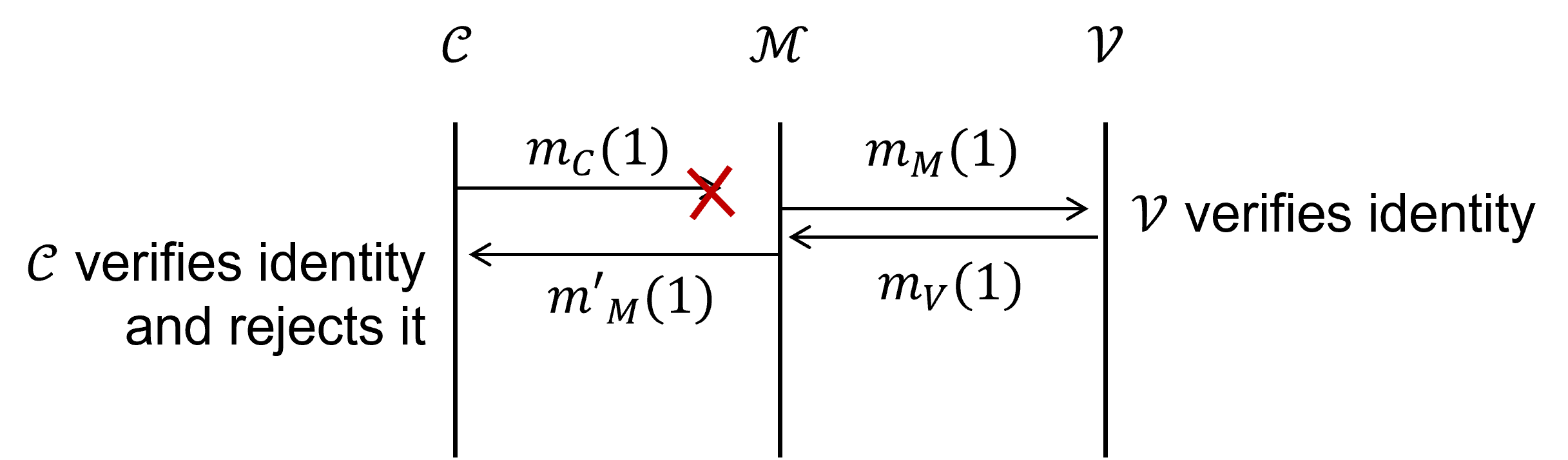}\\
  (a)\\
  \includegraphics[width=0.9\columnwidth]{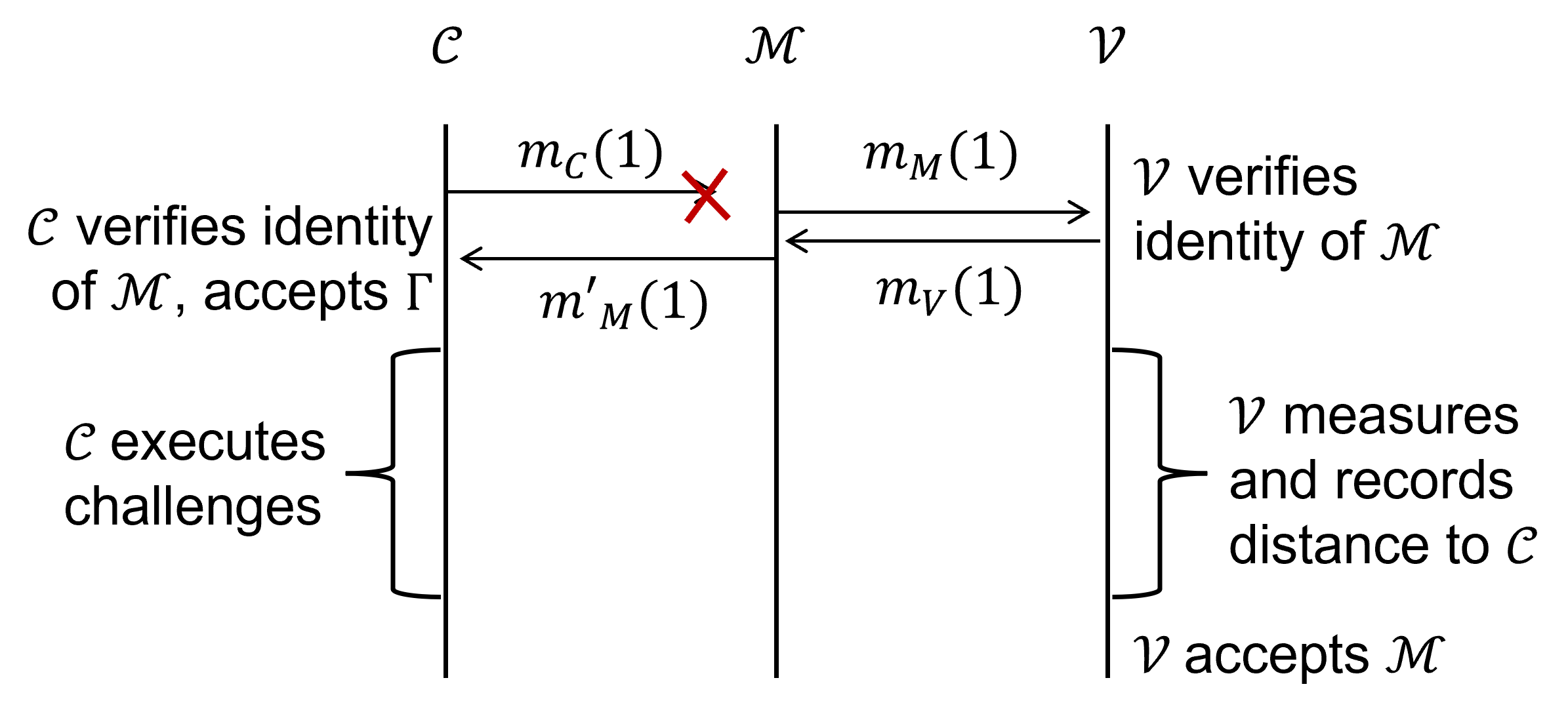}\\
  (b)
\vspace{-0.15in}
\end{tabular}
\caption{A MiTM attack. (a) $ID_V$ and $pk_V$ are known to $\mathcal{C}$, (b) the verifier is unknown to $\mathcal{C}.$}
\label{fig:MiTM}
\vspace{-0.2in}
\end{figure}

{\bf Joining a pre-specified platoon.} Let the candidate target a specific platoon identified by verifier with $(ID_V, pk_V, cert_V)$. The steps of a  MiTM attack are shown in Fig.~\ref{fig:MiTM}(a). The candidate initializes the protocol by sending a join request message $m_C(1)$ to $\mathcal{V}.$ The request $m_C(1)$ contains the $ID_C$ and $ID_V$, signed with the candidate's private key. The adversary can attempt to initiate parallel sessions by eliminating $m_C(1)$ (e.g., via jamming) and injecting his own request to join $\mathcal{V}.$
\[
m_M(1) \leftarrow ID_V,ID_M, pk_M, cert_M, sig_{sk_M}(\text{REQ}, ID_M, ID_V).
\]
Upon receiving $m_M(1)$, the verifier validates the digital identity of $\mathcal{M}$ and challenges $\mathcal{M}$ with $\Gamma$. Because the adversary is not following the platoon, the only chance to successfully complete the MiTM attack is for the valid candidate to execute the physical challenges $\Gamma.$ The adversary can attempt to respond to $\mathcal{C}$'s initial message $m_C(1)$ by sending 
\[
m'_M(1) \leftarrow ID_C, E_{pk_C}[sig_{sk_M}(\Gamma, ID_M, ID_C, t_0), \Gamma, ID_M, ID_C, t_0],
\]
containing the same set of physical challenges $\Gamma$, provided by $\mathcal{V}$ to $\mathcal{M}.$ However, $\mathcal{C}$ will abort the joining process because the reply is signed by $\mathcal{M}$ and not $\mathcal{V}.$ At this stage, the adversary's MiTM attack fails because $\mathcal{C}$ can only accept challenges signed by $\mathcal{V}.$

\textbf{Joining an arbitrary platoon.} When the candidate opportunistically tries to join a platoon, he may not know the verifier's identity. Consider a candidate following $\mathcal{V}$ at $d_{ref}$ but being unaware of the identity of $\mathcal{V}.$  The adversary can then launch a MiTM attack where he spoofs $\mathcal{V}$. The steps of the MitM attack are shown in Fig.~\ref{fig:MiTM}(b). 
The candidate $\mathcal{C}$ initiates a platoon join request by sending message
\[
m_C(1) \leftarrow \text{REQ}, ID_C, pk_C, cert_C.
\]
Note that the join is not directed to a specific verifier (alternatively, the candidate may respond to a probe from nearby verifiers, similar to the reception of SSIDs from nearby Wi-Fi networks, but the end result is the same in terms of knowing the identity of $\mathcal{V}$.) The adversary $\mathcal{M}$ corrupts $m_C(1)$ (e.g., via jamming) to prevent $\mathcal{V}$ from receiving it and initiates his own session with $\mathcal{V}$ by sending 
\[
m_M(1) \leftarrow ID_V,ID_M, pk_M, cert_M, sig_{sk_M}(\text{REQ}, ID_M, ID_V)
\]
to the verifier. The verifier responds to $\mathcal{M}$ with
\[
m_V(1) \leftarrow ID_M, E_{pk_M}[sig_{sk_V}(\Gamma, ID_V, ID_M, t_0), \Gamma, ID_V, ID_M, t_0],
\]
challenging $\mathcal{M}$ with $\Gamma.$ The adversary responds to $m_C(1)$  pretending to be a verifier and challenges the valid candidate with the same physical challenges $\Gamma$ 
\[
\resizebox{.95\hsize}{!}{$m'_M(1) \leftarrow ID_C, E_{pk_C}[sig_{sk_M}(\Gamma, ID_M, ID_C, t_0), \Gamma, ID_M, ID_C, t_0]$}
\]
The candidate executes the perturbations in $\Gamma$ leading $\mathcal{V}$ to admit $\mathcal{M}$, even though $\mathcal{M}$ {\em does not follow $\mathcal{V}$ within distance $d_{ref}.$}

\textbf{Resistance to MiTM attacks.} Preventing the MiTM attack when the identity of the verifier is not known to the valid candidate is a challenging problem. Without any means to authenticate the intended verifier, impersonation is possible. We emphasize the required sophistication to launch such an advanced attack. The adversary must intercept the request of a valid candidate who opportunistically seeks to join a platoon while being unaware of the identity of the verifier representing that platoon. 

Whereas the {\em Wiggle} protocol does not prevent this kind of attack, we present some possible directions to remedy it. One potential solution is to use highly-directional antennas on the candidate and the verifier (e.g., at mmWave frequencies). By pointing the antenna of the candidate in the forward direction within the travelling lane and the verifier's antenna in the backward direction, the adversary has a limited opportunity to launch the MiTM attack. Another candidate solution is to employ a single-receiver transmission localization system that can pinpoint the location of the transmitter (e.g., \cite{soltanaghaei2018multipath}). Although such systems have been demonstrated to be highly accurate, they have been considered under static scenarios rather than high velocity mobile setups. An alternative direction would be to exploit the Doppler shift between the candidate and the verifier to detect messages injected by the adversary. Given the applicability of this attack only to opportunistic platooning scenarios, we leave these directions as future work.

\section{Evaluation}
In this section, we evaluate the security and performance of the \textit{Wiggle} protocol.  All platooning experiments were performed in the Plexe simulation environment \cite{segata2022multi}, which  is a cooperative driving framework permitting the realistic simulation of platooning systems. It features realistic vehicle dynamics and several cruise control models, enabling the analysis of mixed scenarios in traffic.

\begin{figure*}%
\centering
\begin{tabular}{ccc}
  \includegraphics[width=0.65\columnwidth]{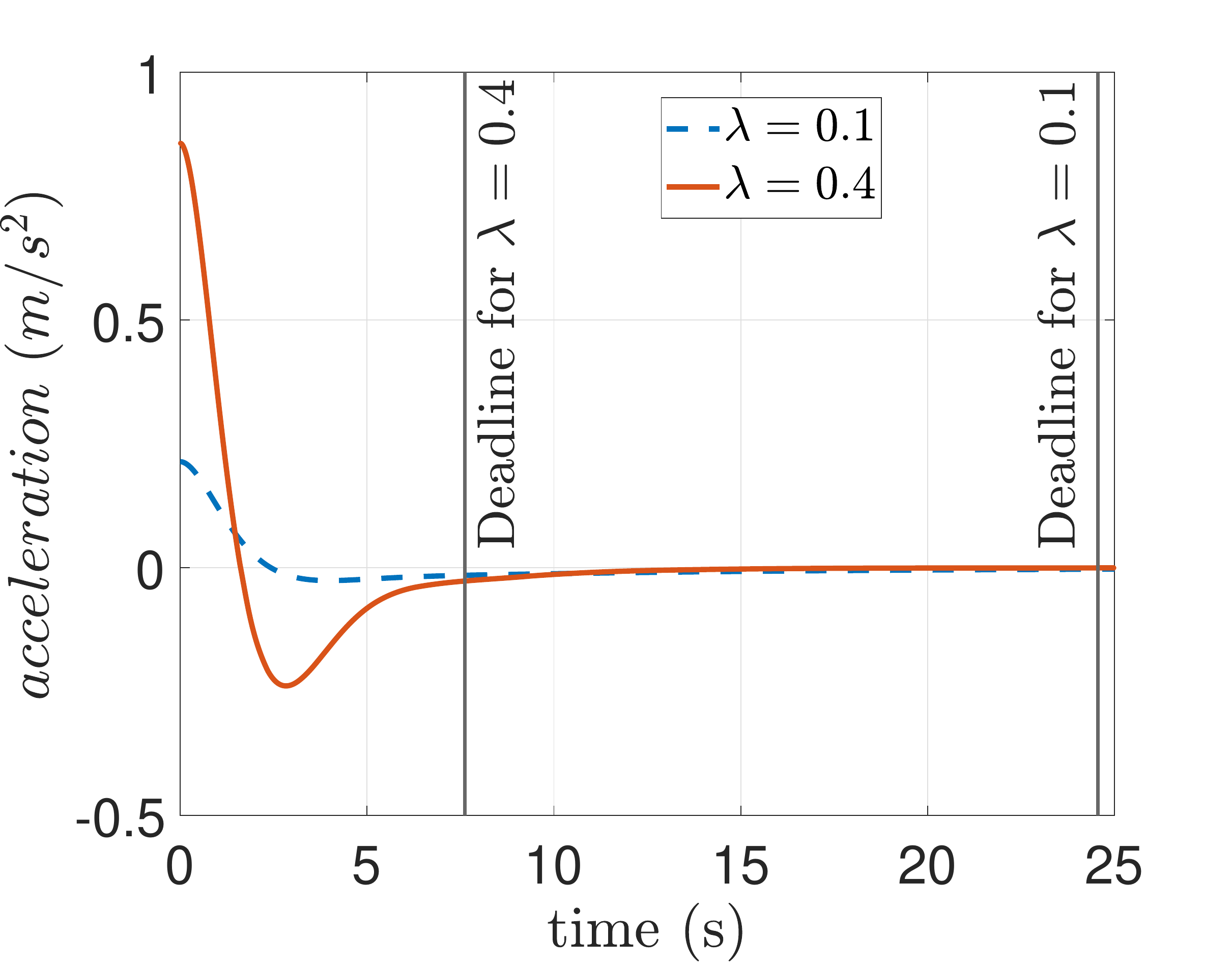} &
  \includegraphics[width=0.65\columnwidth]{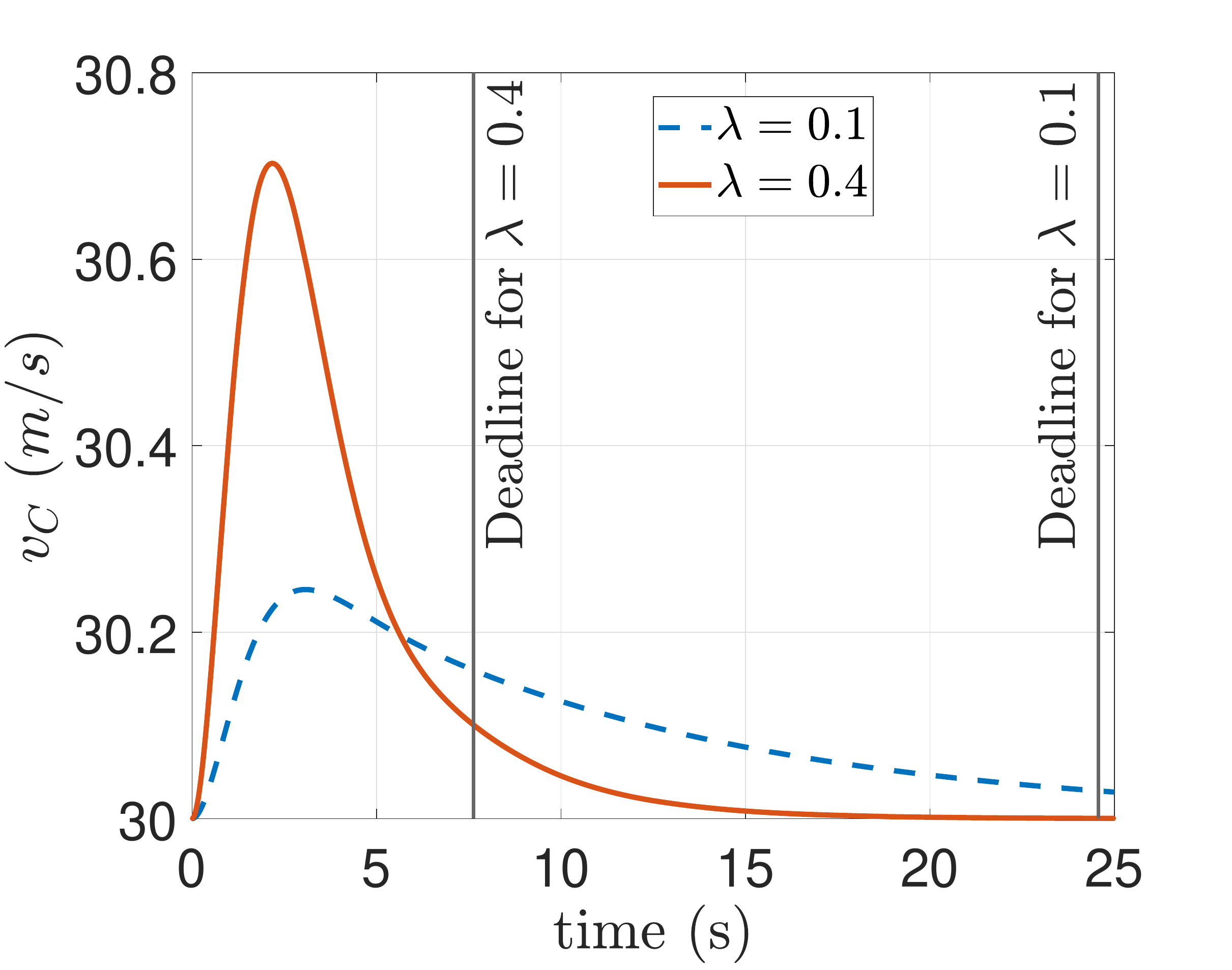} & 
  \includegraphics[width=0.65\columnwidth]{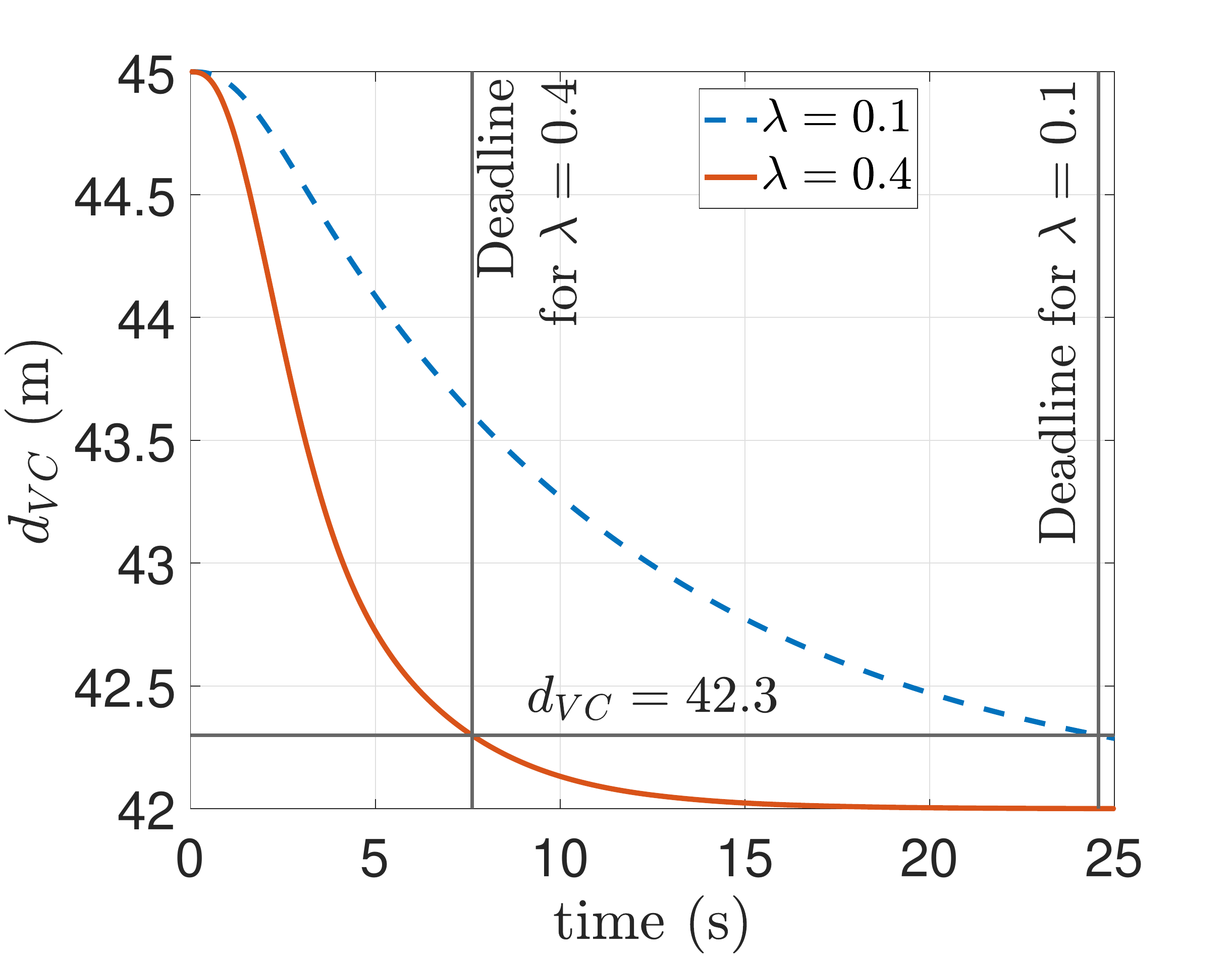} \\
  (a) acceleration,& (b) velocity & (c) following distance
\vspace{-0.15in}
\end{tabular}
\caption{The acceleration, velocity, and distance of $\mathcal{C}$ to reach checkpoint $d=42$m from $d_{ref}=45$m, when $v_C=30$m/s and $\lambda$ is set to 0.1 and 0.4.}
\label{fig:acc_control}
\vspace{-0.2in}
\end{figure*}

\begin{table}[t]
\centering
\caption{Simulation Parameters}
\vspace{-0.1in}
    \begin{tabular}{ll}
    \hline
    Parameter & Value \\
    \hline
    Velocity of $\mathcal{V}$ ($v_V$) and $\mathcal{C}$ ($v_C$) & 30m/s \\
    Following distance ($d_{ref}$) & $1.5 \cdot v_C$ (45m) \\
    Checkpoint range & $1\cdot v_C - 2\cdot v_C$ (30m $-$ 60m)\\
    \# of checkpoints in range ($M$) & 51 \\
    Update step of ACC ($\Delta_t$) & 0.1s \\
    ACC parameter $\lambda$ & 0.4 \\
    Checkpoint error tolerance ($\gamma$) &  0.3m \\
    \hline
    \end{tabular}
\label{tab:simulation_Parameters}
\vspace{-0.15in}
\end{table}

\subsection{Performance of Wiggle}
\label{perEval}

We first evaluated the performance of {\em Wiggle} as a function of the different protocol parameters. In our simulation, a verifier $\mathcal{V}$ was followed by a candidate $\mathcal{C}$ in a freeway environment. The candidate applied the ACC model presented in Section~\ref{sec:ACCcontrol} to control its following distance from the verifier. The simulation parameters  are listed in Table. \ref{tab:simulation_Parameters}. Initially, the verifier and the candidate were platooning at a speed of 30m/s (108Km/h) without any interfering traffic. The following distance $d_{ref}$ was set to 45m, which corresponds to a 1.5sec time gap. The verifier challenged the candidate to perform perturbations within a range of 30m (1sec) around the following distance by generating physical challenges at random. 

{\bf Verification time:} The performance of {\em Wiggle} was evaluated in terms of the delay until a candidate is admitted to the platoon. Intuitively, delay is a function of the ACC parameters  and the number of physical challenges.

{\bf Studying the impact of the ACC.} The ACC parameters control the deadline for reaching each checkpoint. Parameter $\lambda$, in particular, regulates the vehicle acceleration as a function of the distance to the checkpoint. Figure~\ref{fig:acc_control} shows the candidate's acceleration, velocity, and following distance as a function of time, when the checkpoint is 3m away from $d_{ref}.$ From Fig.~\ref{fig:acc_control}(a), we observe that the initial applied acceleration is gradually decreased, and then the vehicle breaks until the checkpoint is reached. The speed differential hardly exceeds 0.6m/sec (2Km/h), indicating an almost imperceptible transition to the checkpoint. We further observe that when $\lambda$ is decreased to 0.1, the acceleration and velocity differential decrease at the expense of longer delay until the checkpoint is reached. In the remaining of our simulations, we set $\lambda=0.4.$

\begin{figure}[t]
\centering
\setlength{\tabcolsep}{-3pt}
\begin{tabular}{cc}
\includegraphics[width=0.55\columnwidth]{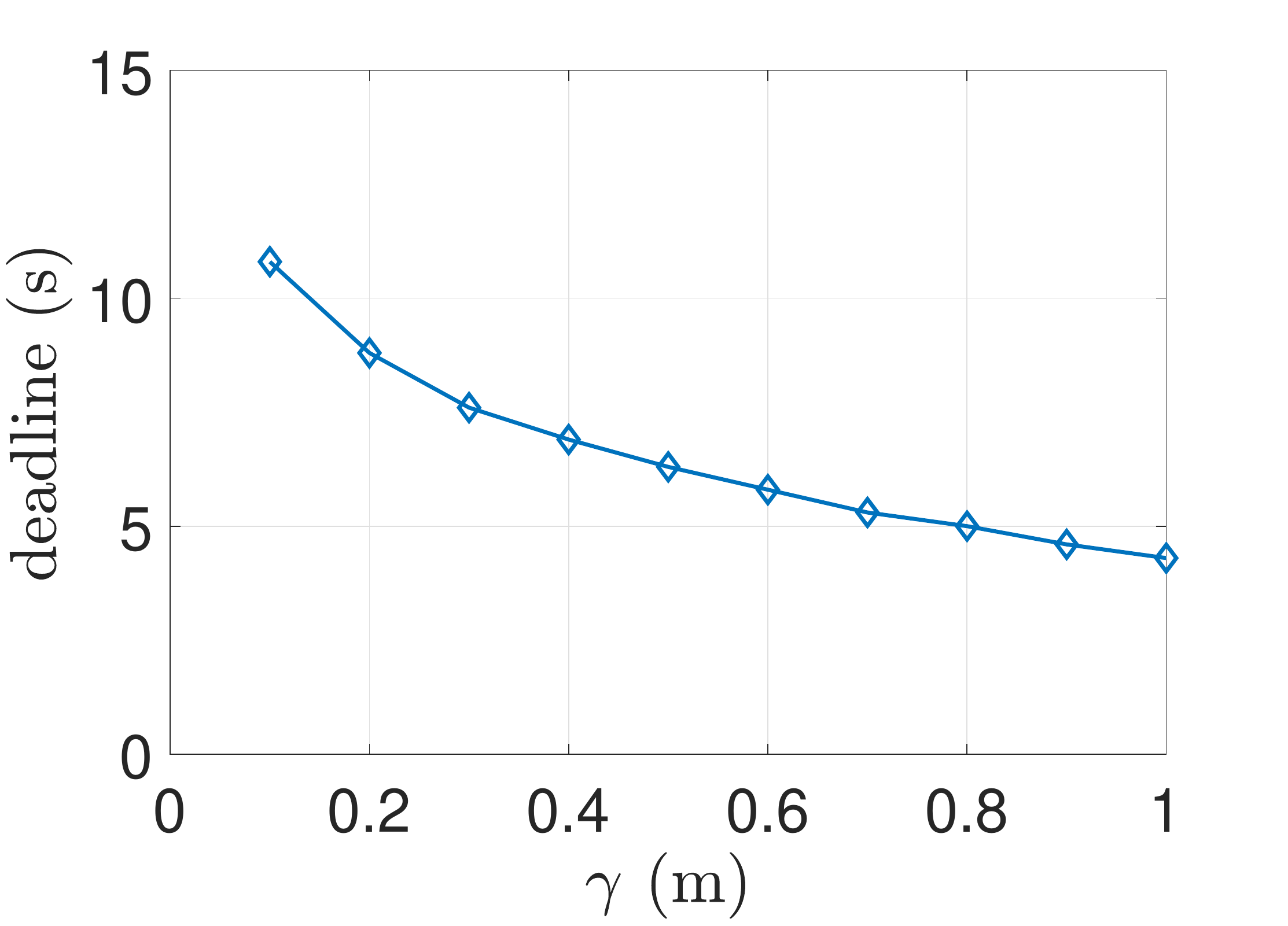}&
\includegraphics[width=0.55\columnwidth]{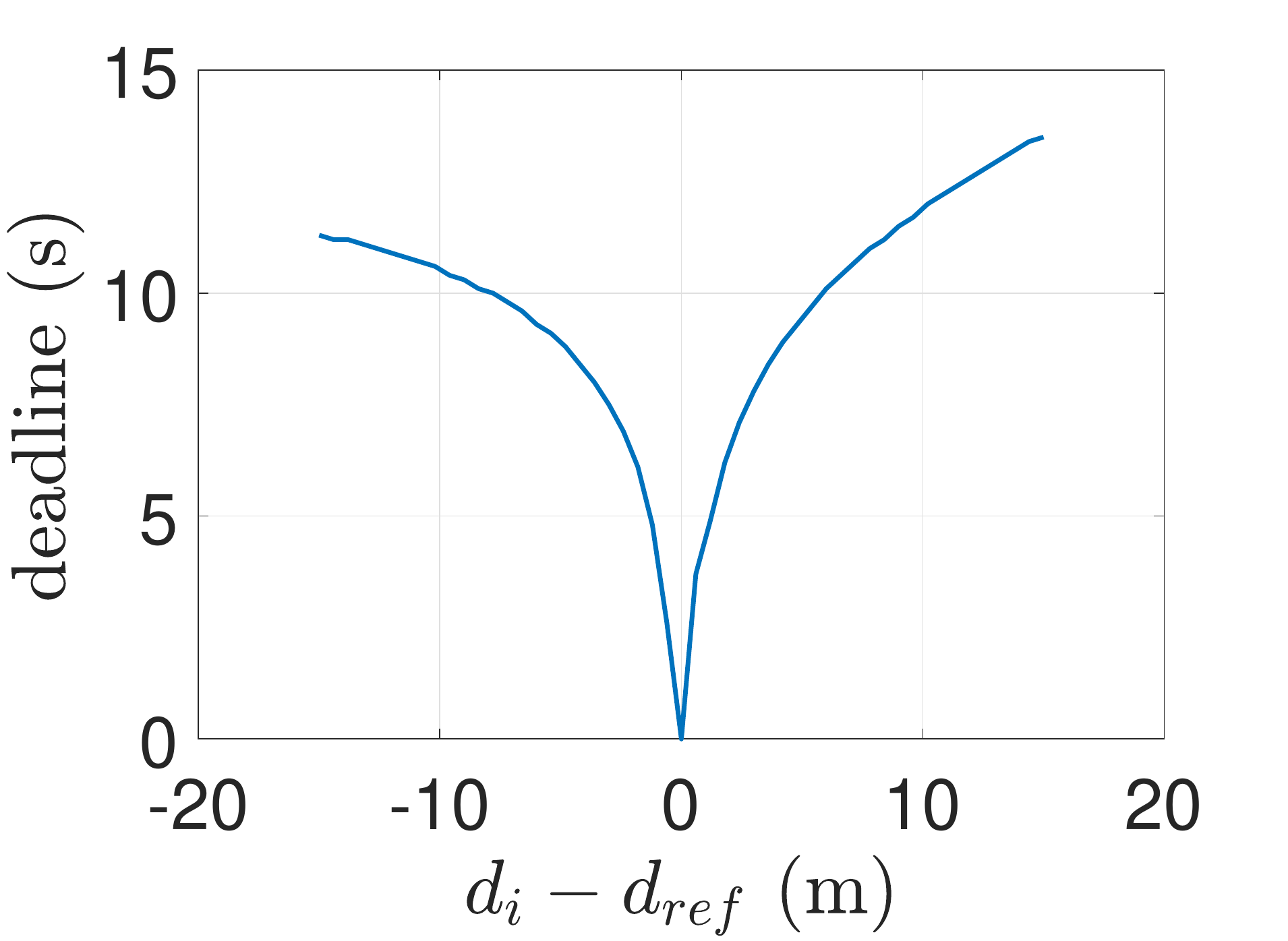}\\
(a) & (b)
\vspace{-0.15in}
\end{tabular}
\caption{(a) The deadline duration as a function of the checkpoint distance threshold $\gamma$ for a checkpoint 3m away from $d_{ref}$ and (b) the deadline duration as a function of the distance covered by checkpoints, when $\gamma=0.3$m.}
\label{fig:gamma_time}
\vspace{-0.2in}
\end{figure} 

Another important parameter that impacts delay is the distance tolerance $\gamma$ by which the checkpoint must be reached. Figure~\ref{fig:acc_control} shows that the vehicle quickly  converges in the vicinity of the checkpoint and then fine tunes its position to reach the checkpoint. By increasing the distance tolerance, the deadline can be shortened. Figure~\ref{fig:gamma_time}(a) indeed shows that the deadline duration is inversely related to $\gamma$. For our simulations, we selected $\gamma=0.3$m, which is 
close to the typical automotive radar resolution \cite{waldschmidt2021automotive}. 

Finally, in Fig.~\ref{fig:gamma_time}(b), we show the deadline as the function of the distance that the candidate has to cover to reach the checkpoint. We observe that the deadline grows with distance but the relationship is not linear. This is justified by the acceleration model of the ACC model. We also note that the deadlines are not symmetric when the same distance has to be covered forward and backward as slightly different accelerations are applied in each direction.

\begin{figure*}%
\centering
\setlength{\tabcolsep}{-3pt}
\begin{tabular}{ccc}
  \includegraphics[width=0.75\columnwidth]{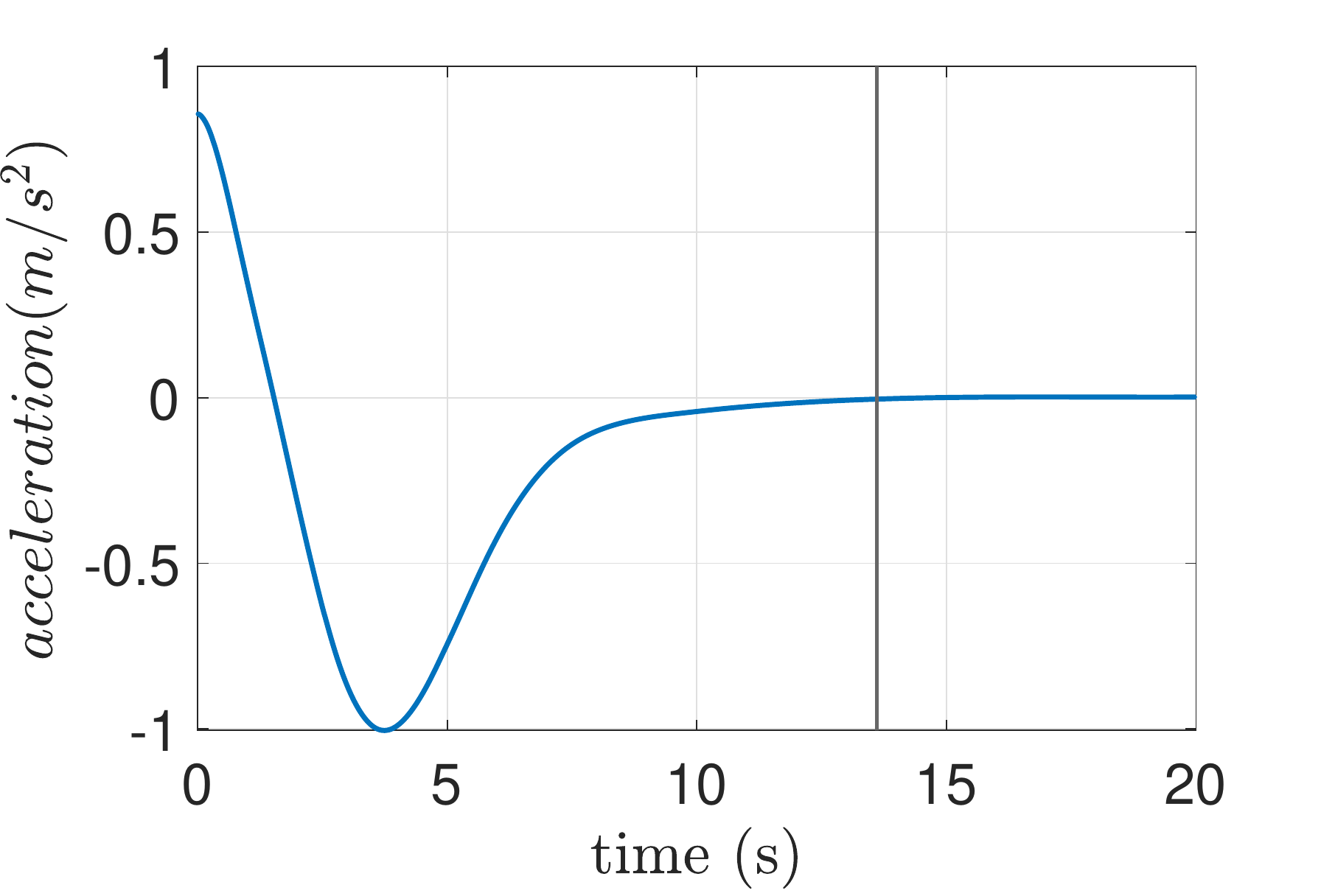} &
  \includegraphics[width=0.75\columnwidth]{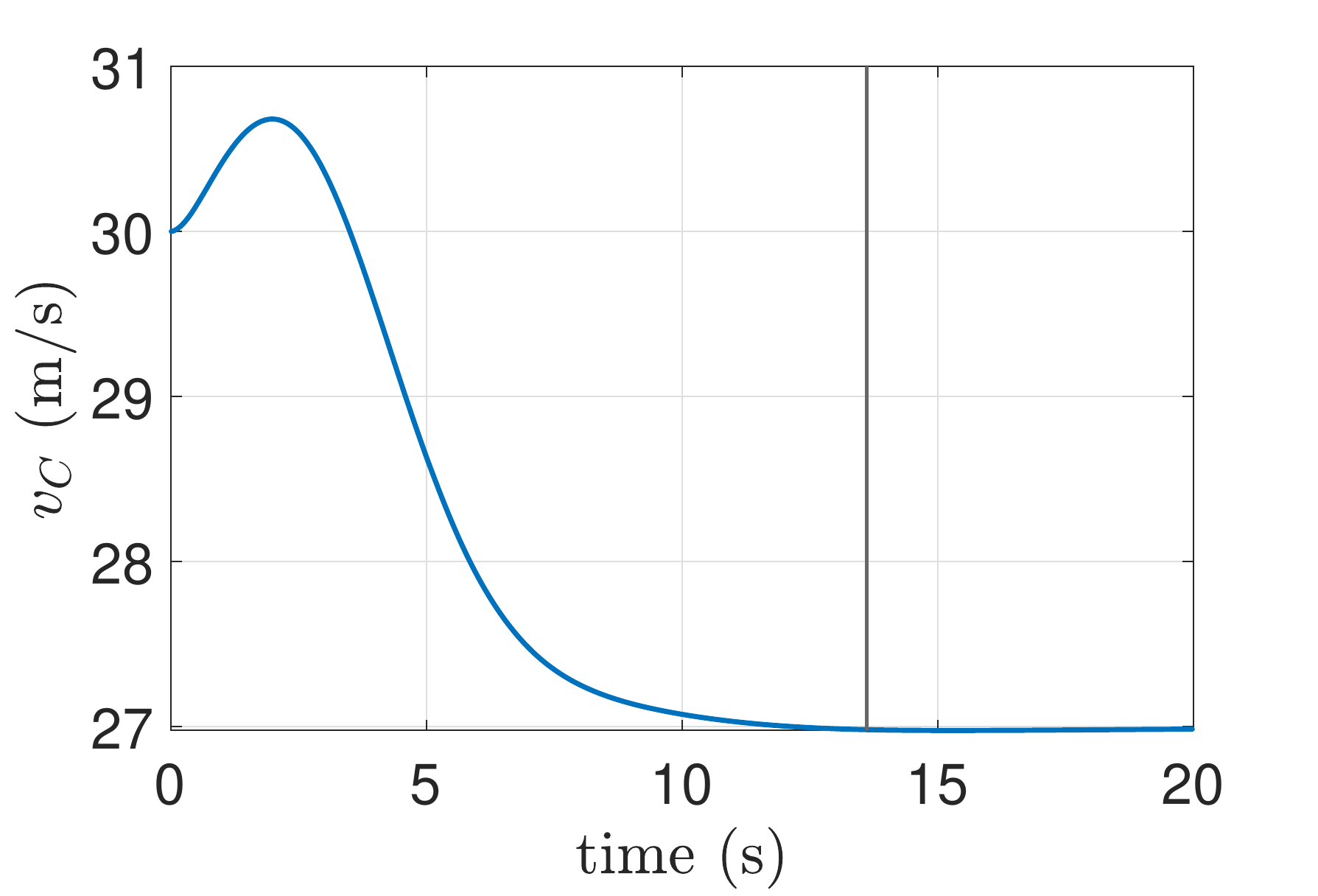} & 
  \includegraphics[width=0.75\columnwidth]{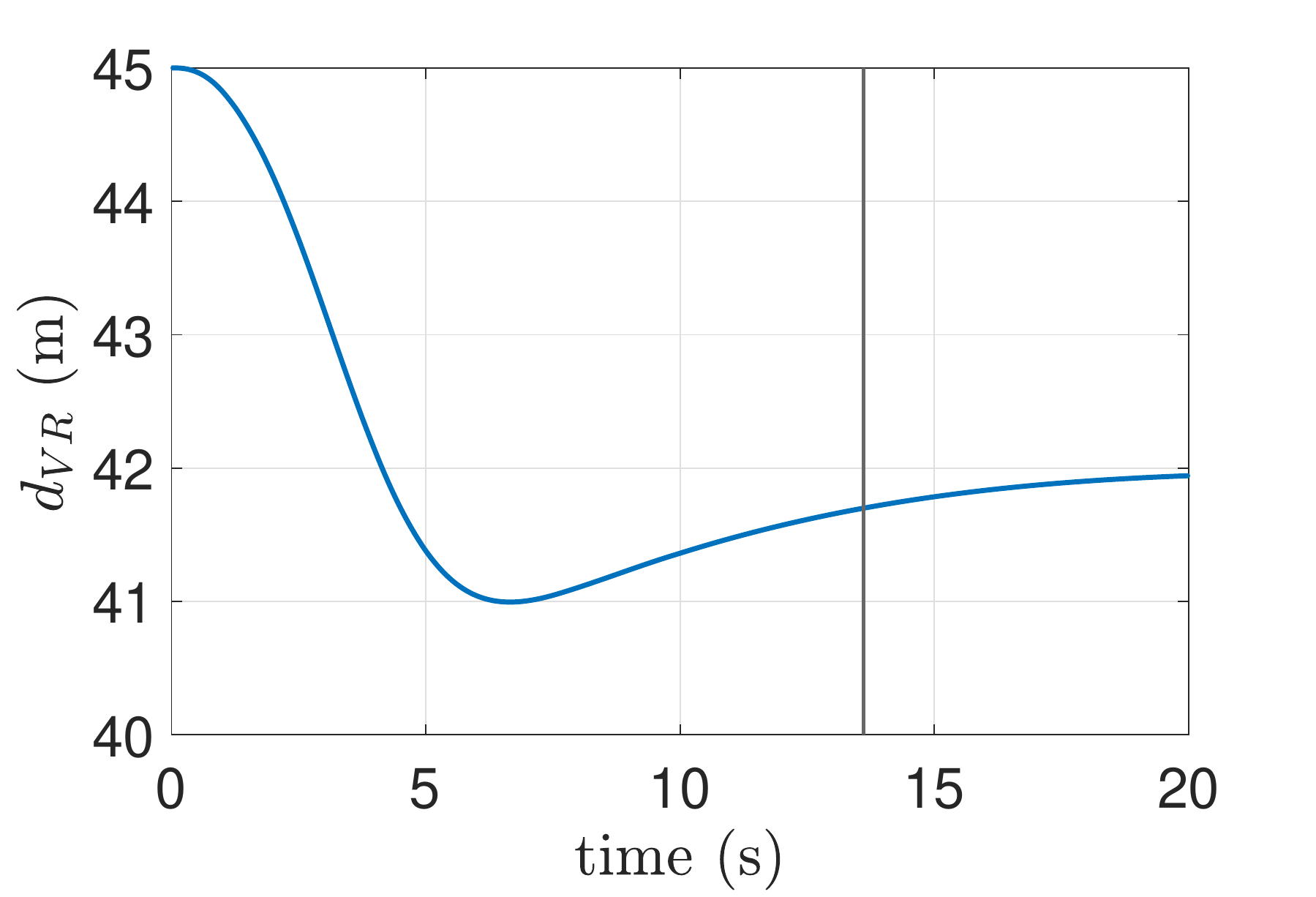} \\
  (a) acceleration & (b) velocity & (c) following distance
\vspace{-0.15in}
\end{tabular}
\caption{The acceleration, velocity, and distance of $\mathcal{C}$ from $d_{ref}=45$m to reach checkpoint $d=42$m, when the velocity of the verifier reduces from $30m/s$ to $27m/s$ during the verification.}
\label{fig:acc_control_delay}
\vspace{-0.2in}
\end{figure*}

\textbf{Impact of traffic.} So far, we have assumed that the verifier moves at constant velocity when the candidate responds to physical challenges. However, traffic may impact the velocity of the verifier and the way that the candidate's ACC approaches a checkpoint. To study this impact, we simulated a vehicle proceeding the verifier traveling at 27m/s. To maintain a safe distance when $\mathcal{V}$ comes upon the slow vehicle, $\mathcal{V}$ reduces its velocity to 27m/s while the candidate is attempting to reach a checkpoint. Figure~\ref{fig:acc_control_delay} shows the acceleration, velocity, and following distance of the candidate as a function of time for a checkpoint that is at 42m from $\mathcal{V}.$ We observe that the time to reach the checkpoint increased from 7.6sec (according to Fig.~\ref{fig:acc_control}) to 13.6sec. Moreover, the candidate actually moved passed the checkpoint before recovering to the checkpoint due to $\mathcal{V}$'s braking. This indicates that a valid candidate will fail the original deadline, if the velocity of the verifier changes. 

There are two approaches to remedy this problem. The first is to ignore any challenges for which the verifier's velocity changes drastically and repeat them when the velocity stabilizes. The second approach is for the verifier to adjust the deadline based on his own velocity. Given the ACC model, the verifier can re-compute the deadline to allow for the candidate to reach the checkpoint. 

\begin{figure}[t]%
\centering
\setlength{\tabcolsep}{-3pt}
\begin{tabular}{cc}
  \includegraphics[width=0.55\columnwidth]{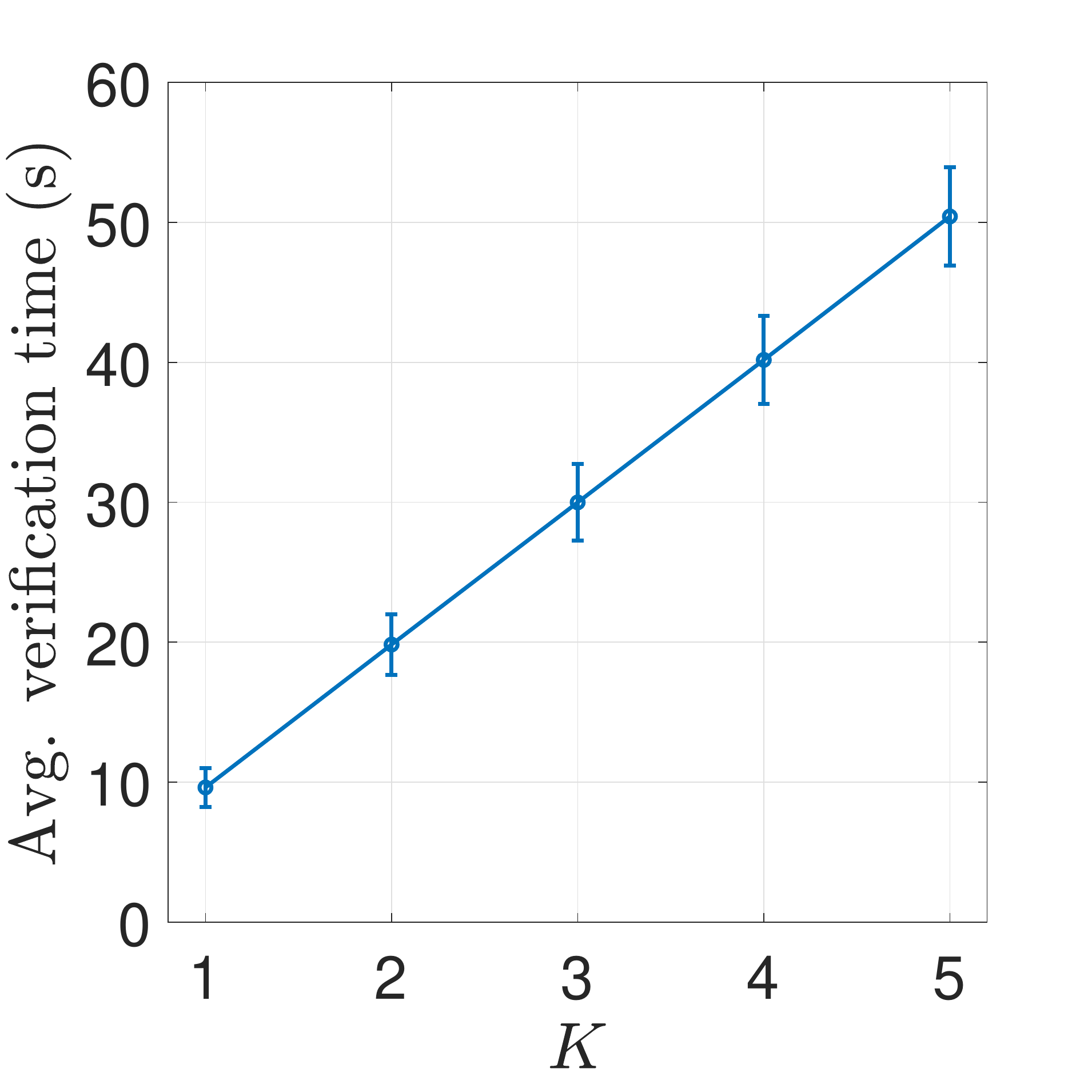} &
  \includegraphics[width=0.55\columnwidth]{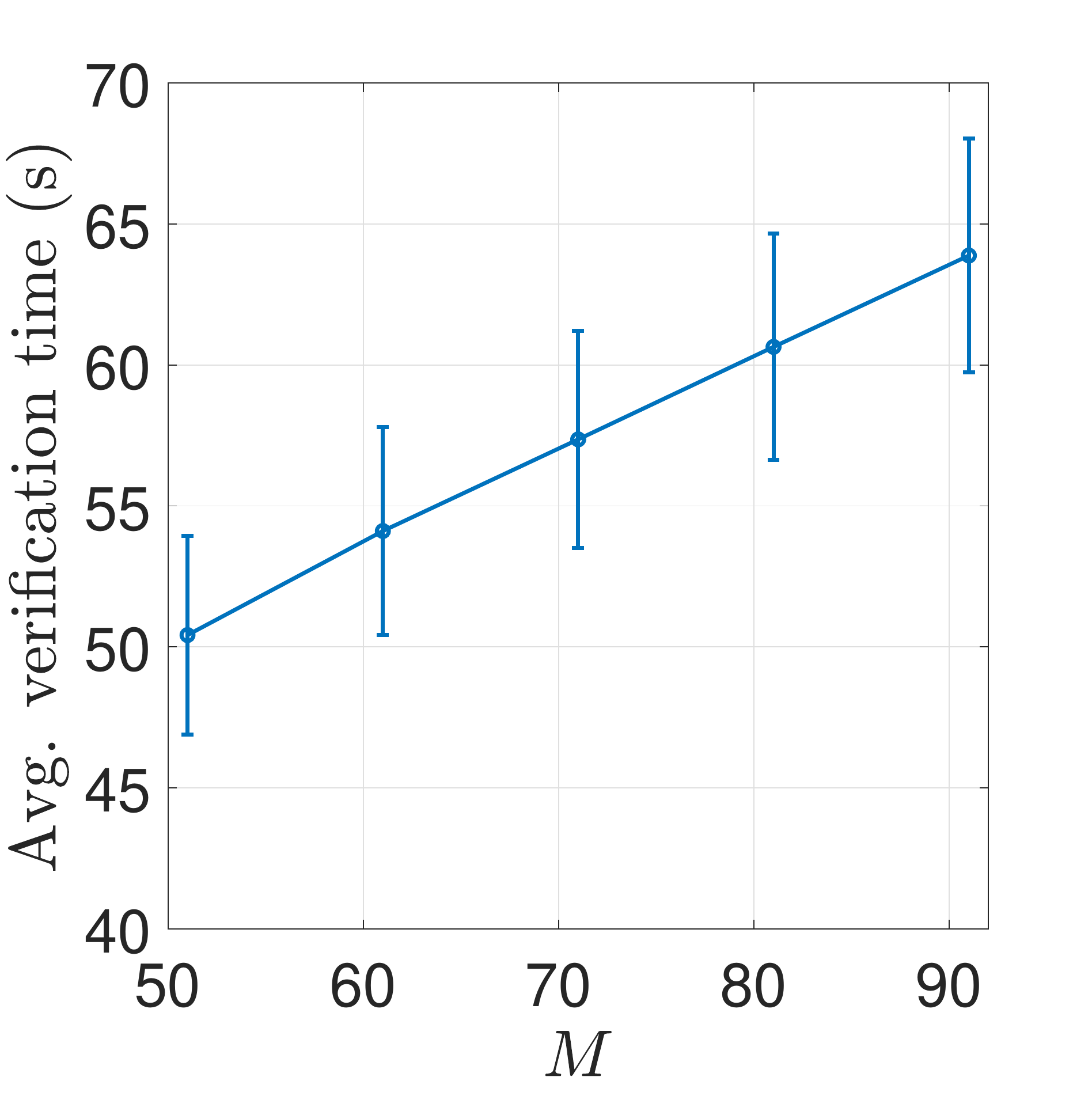}
\vspace{-0.2in}
\end{tabular}
\caption{Verification time as a function of the number of challenges $K$ and available checkpoints $M$.}
\label{fig:avg_time_K_M}
\vspace{-0.2in}
\end{figure}

{\bf Verification time as a function of physical challenges $K$.} The verification time also depends on the number of physical challenges issued by the verifier. Indeed, this relationship is expected to be linear as the verification delay is cumulative with every challenge. Variations are due to the variability of the deadlines for randomly selected checkpoints. To study the impact of $K$, we fixed the checkpoint space to $M=51$ and varied $K$ while executing {\em Wiggle}. Figure~\ref{fig:avg_time_K_M}(a) shows the average verification time and its standard deviation as a function of $K$. We observe the expected linear increase in verification time, with about 10sec overhead per physical challenge. Overall, the verification time is short (less than a minute) relative to the time that the candidate will be platooning with the rest of the platoon.   Figure~\ref{fig:avg_time_K_M}(b) shows the average verification time as a function of the number of available checkpoints $M,$ when $K=5$. As the range of motion of the candidate expands, the verification time increases due to the longer average distance to reach each checkpoint.

\subsection{Security of Wiggle}

\begin{figure}[t]%
\centering
\setlength{\tabcolsep}{-3pt}
\begin{tabular}{cc}
  \includegraphics[width=0.55\columnwidth]{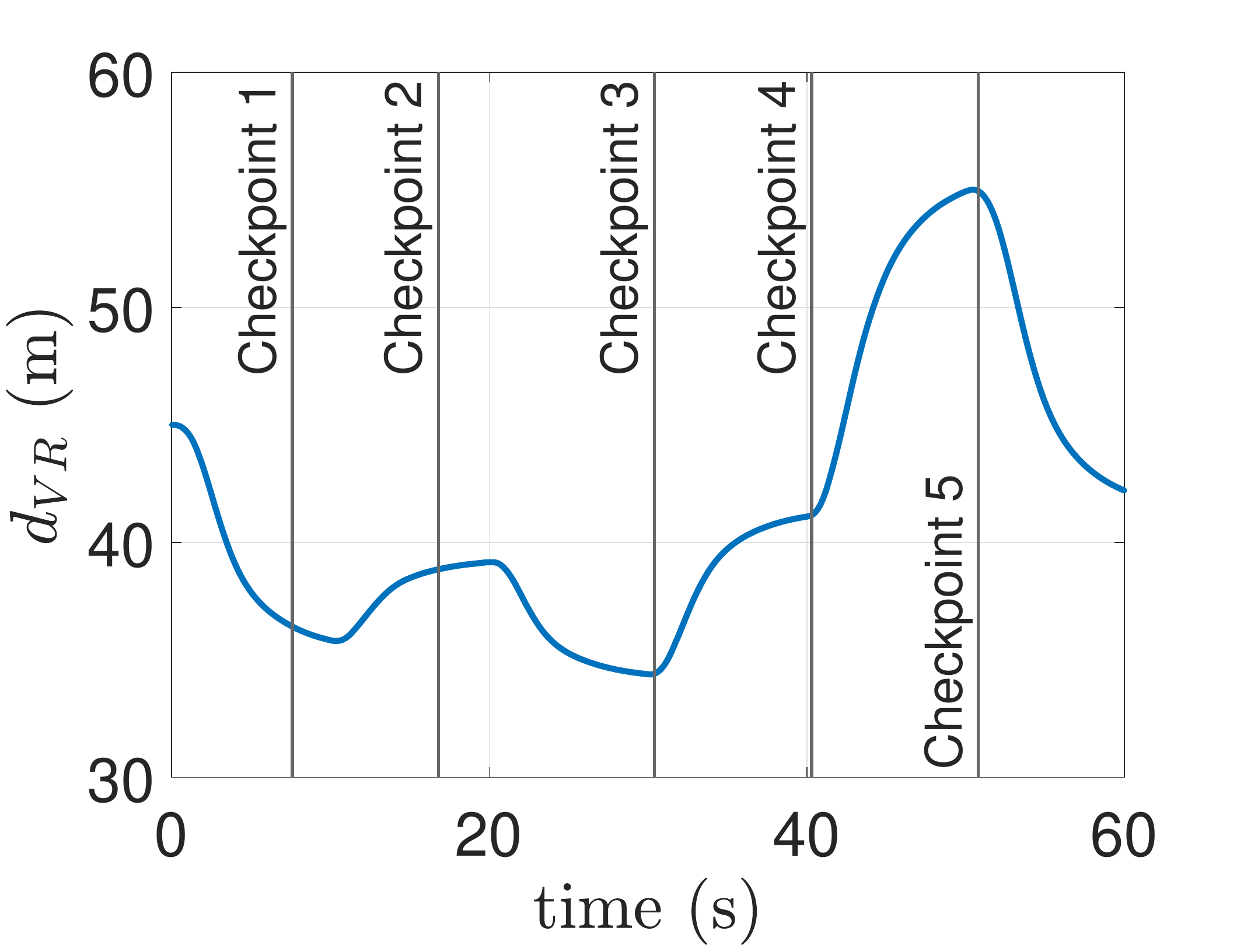} &
  \includegraphics[width=0.55\columnwidth]{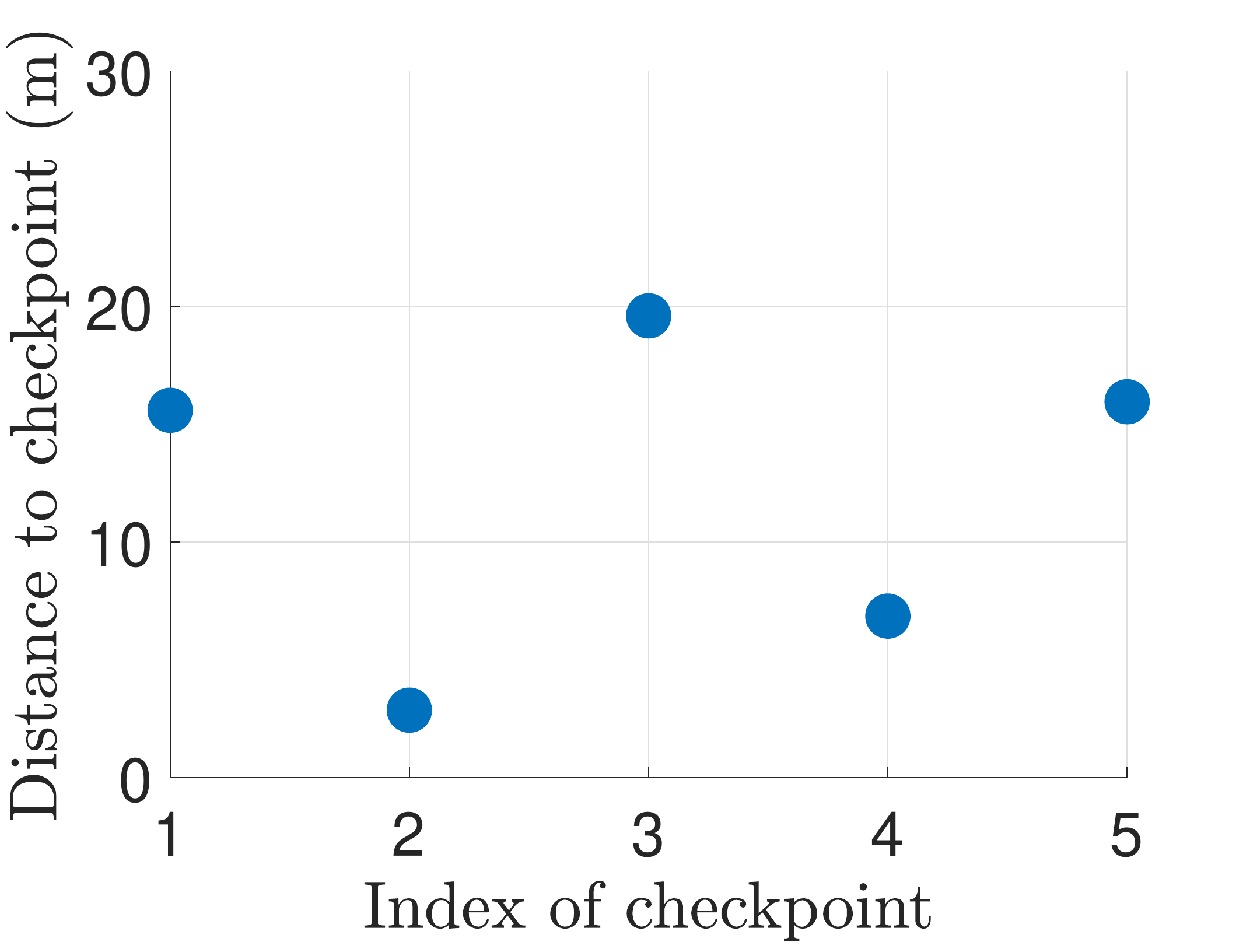} \\ 
  (a) & (b) 
\vspace{-0.15in}
\end{tabular}
\caption{(a) The distance between  $\mathcal{R}$ and $\mathcal{V}$ as a function over five challenges, (b) the distance difference between the vehicle $\mathcal{R}$ and the checkpoints at each deadline.}
\label{fig:R_dis_error}
\vspace{-0.2in}
\end{figure}

In Section \ref{ssec:remote}, we showed that a remote adversary is unable to pass the PoF verification without performing the physical challenges. The only chance for the adversary is that some independent vehicle $\mathcal{R}$ follows the verifier at the platooning distance. We evaluated the probability that $\mathcal{M}$ passes verification due to $\mathcal{R}$'s motion, as stated in Proposition 1. We simulated a vehicle $\mathcal{R}$ following a verifier traveling at 30m/sec. The vehicle $\mathcal{R}$ executed a random walk within the checkpoint range (30m - 60m from the verifier) with a step size of 0.3m (i.e., $N$=100 Markov states). The verifier continuously issued physical challenges with a distance tolerance of $\gamma=0.3$m. Figure~\ref{fig:R_dis_error}(a) shows an instance of $\mathcal{R}$'s following distance to $\mathcal{V}$ as a function of time for five checkpoints. Figure~\ref{fig:R_dis_error}(b) shows the distance of $\mathcal{R}$ from each checkpoint at the time of the deadline. We observe that $\mathcal{R}$ is often at a location far away from the respective checkpoint, since it does not try to reach it intentionally. 

\begin{figure}[t]%
\centering
\setlength{\tabcolsep}{-3pt}
\begin{tabular}{cc}
  \includegraphics[width=0.58\columnwidth]{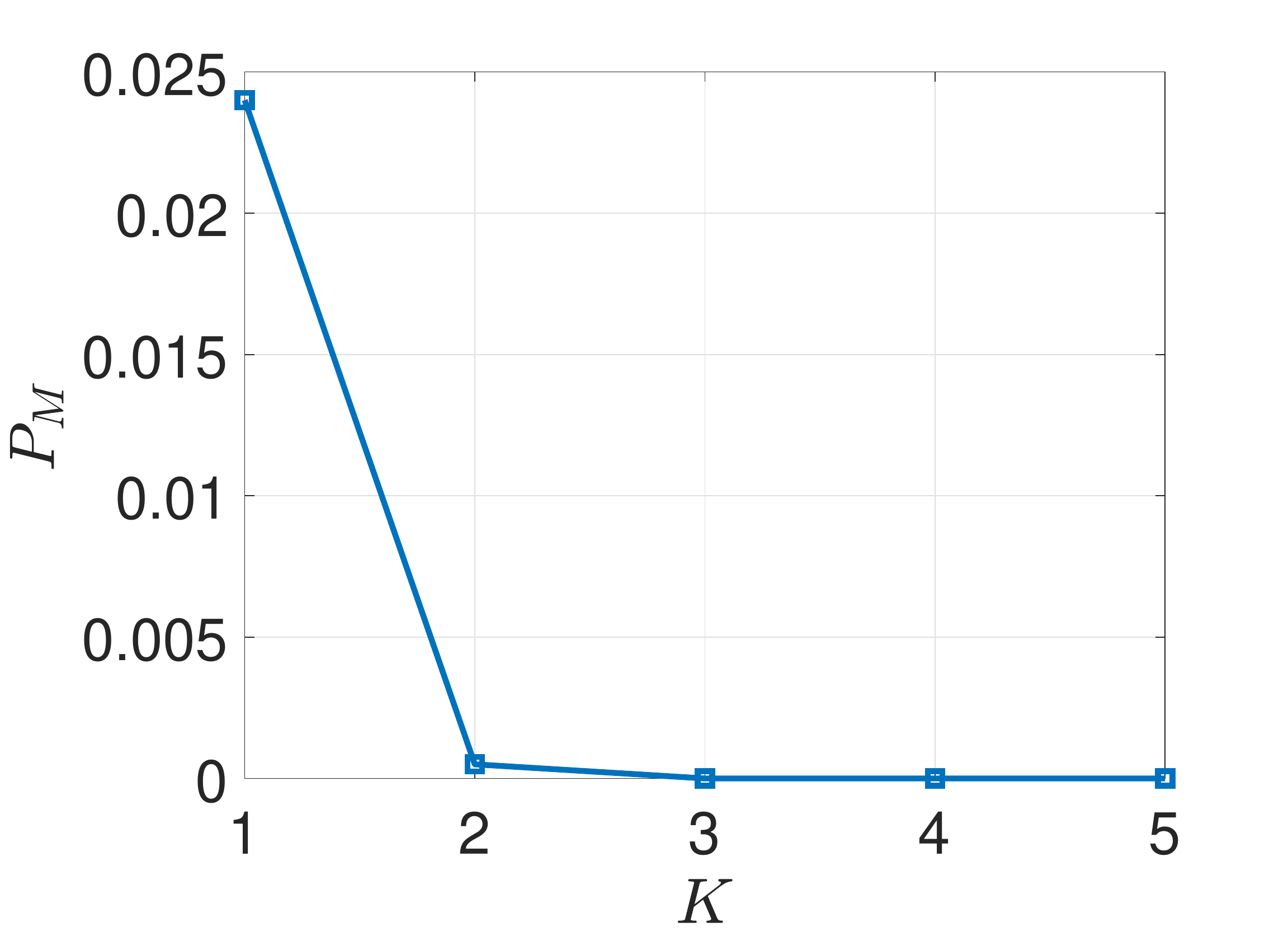} &
  \includegraphics[width=0.52\columnwidth]{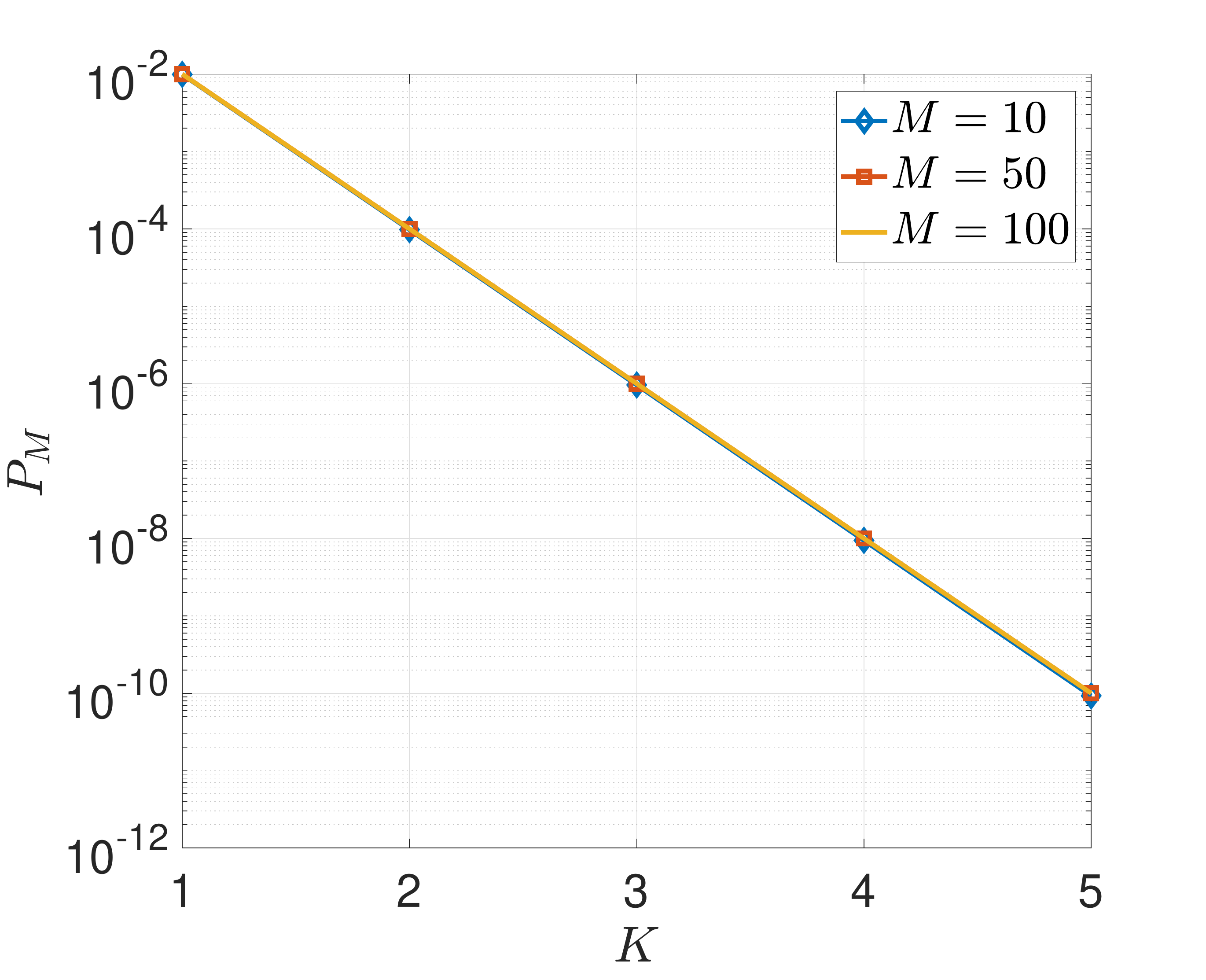} \\
  (a)  simulation & (b) numerical
\vspace{-0.15in}
\end{tabular}
\caption{The simulated and numerical passing probability $\mathbf{P}_M$ as a function of the number of challenges $K$.}
\label{fig:PM}
\vspace{-0.2in}
\end{figure}

This is further verified in Fig.~\ref{fig:PM}(a) that shows $\mathcal{M}$'s passing rate as a function of the number of physical challenges, calculated over 2,000 challenges. Note that after $K=2$, $\mathcal{M}$ did not pass any of the PoFs $(P_M =0)$. For comparison, we also provide $P_{M}$ when calculated numerically using Proposition 1. A few physical challenges are sufficient to drive the probability of success to very low values. Note that the checkpoint space cardinality $M$ does not affect $P_M$. This is because  $\mathcal{R}$ must reach one specific checkpoint by the deadline. It is fairly straightforward to show that under a random walk, this probability follows the uniform distribution (with slightly higher probabilities for the two boundaries). Therefore, regardless of $M,$ $P_M$ is approximately equal to $\left(1 \slash N\right)^K$, as it is also observed by the numerical results in Fig.~\ref{fig:avg_time_K_M}(b). 

\section{Related Work}

\textbf{Verification of Platooning.} Several prior works have considered the problem of access control for platoon admission.  \cite{han2017convoy, xu2021pof, vaas2018get, juuti2017stash}. Han {\em et  al.} introduced {\em Convoy,} a platoon admission method that relies on physical context \cite{han2017convoy}. Convoy exploited the correlation between the vertical acceleration recorded at the candidate and the verifier due to the variations of the road surface. However, this approach is vulnerable to record and replay attacks since the road surface condition changes slowly over time. Moreover, it cannot precisely determine the following distance. Vaas $et\ al.$ \cite{vaas2018get} and Juuti $et\ al.$ \cite{juuti2017stash} used the driving trajectory as a proof for platoon membership. A candidate recorded its own trajectory and reported it to the verifier as a proof of platooning. However, the platoon trajectory can be known a priori or can be monitored from a distance. 

Xu $et\ al.$  formally defined the PoF concept for vehicle platoons \cite{xu2021pof}. They proposed a platooning verification method that leverages the large-scale fading effect of ambient cellular signals to prove that a candidate is co-travelling with a verifier. The main advantage is that RF signals are highly-dynamic in space and time, and therefore this method is resistant to pre-recording attacks. However, the relative prover-verifier position cannot be determined. Moreover, the method cannot prove that the two parties platoon on the same lane and precisely estimate the following distance. The {\em Wiggle} protocol addresses all these shortcomings by proving the relative vehicle ordering, the exact following distance, and achieving lane verification while maintaining resistance to pre-recording attacks.

 \textbf{Physical challenge-response protocols.}
The idea of a physical challenge-response has been used to achieve various security properties. Shoukry $et\ al.$ proposed PyCRA, an authentication scheme for protecting sensors from physical attacks  \cite{shoukry2015pycra}. In PyCRA, random but known physical probe signals are injected to the environment to validate the correct operation of sensors and prevent analog injection attacks. Although PyCRA can be used to verify sensors such as radar distance estimators, it requires radar downtime and may not be suitable for safety-critical applications.

 Dutta $et\ al.$ \cite{dutta2017estimation} improved the accuracy of PyCRA for distance sensors by minimizing the distance error between the measured and actual distance using a recursive least square method. However, their approach requires the actual distance to be known a priori, which is not realistic. Kapoor $et\ al.$  \cite{kapoor2018detecting} utilized the spatio-temporal correlation of transmissions from MIMO antennas to address the limitations of the prior systems, creating a spatio-temporal physical challenge-response system \cite{kapoor2018detecting}. The automotive radar does not need to be turned off while verification of its accuracy is performed. We note that these works are orthogonal to ours, as verifying the distance to the following vehicle is insufficient to bind it to its digital identity. However, they are useful in securing the sensing modality that is used by {\em Wiggle} to verify the physical challenges.

\section{Conclusion}

We proposed {\em Wiggle}, a physical challenge-response protocol for controlling physical access to a platoon. {\em Wiggle} uses random perturbations of the following distance to bind the digital identity of a candidate to his claimed trajectory. We showed that {\em Wiggle} can verify the following distance of the candidate, the relative positioning of the candidate and the verifier, the candidate's lane, and provide resistance to pre-recording attacks. We evaluated the performance and security of {\em Wiggle} in the  Plexe simulator and showed that a PoF verification lasts less than a minute while inducing almost imperceptible changes to the vehicle's velocity.

\bibliographystyle{ACM-Reference-Format}
\bibliography{wiggle}


\begin{thebibliography}{23}


\ifx \showCODEN    \undefined \def \showCODEN     #1{\unskip}     \fi
\ifx \showDOI      \undefined \def \showDOI       #1{#1}\fi
\ifx \showISBNx    \undefined \def \showISBNx     #1{\unskip}     \fi
\ifx \showISBNxiii \undefined \def \showISBNxiii  #1{\unskip}     \fi
\ifx \showISSN     \undefined \def \showISSN      #1{\unskip}     \fi
\ifx \showLCCN     \undefined \def \showLCCN      #1{\unskip}     \fi
\ifx \shownote     \undefined \def \shownote      #1{#1}          \fi
\ifx \showarticletitle \undefined \def \showarticletitle #1{#1}   \fi
\ifx \showURL      \undefined \def \showURL       {\relax}        \fi
\providecommand\bibfield[2]{#2}
\providecommand\bibinfo[2]{#2}
\providecommand\natexlab[1]{#1}
\providecommand\showeprint[2][]{arXiv:#2}

\bibitem[sec(2020)]%
        {secureV2X}
 \bibinfo{year}{2020}\natexlab{}.
\newblock \bibinfo{title}{3rd Generation Partnership Project;Technical
  Specification Group Services and System Aspects;Security aspect for LTE
  support of Vehicle-to-Everything (V2X) services Rel-16, V16.0.0}.
\newblock
\newblock


\bibitem[IEE(2020)]%
        {IEEE:WAVE}
 \bibinfo{year}{2020}\natexlab{}.
\newblock \bibinfo{title}{IEEE Standard for Wireless Access in Vehicular
  Environments (WAVE)--Certificate Management Interfaces for End Entities}.
\newblock
\newblock


\bibitem[Alam et~al\mbox{.}(2015)]%
        {alam2015heavy}
\bibfield{author}{\bibinfo{person}{Assad Alam}, \bibinfo{person}{Bart
  Besselink}, \bibinfo{person}{Valerio Turri}, \bibinfo{person}{Jonas
  M{\aa}rtensson}, {and} \bibinfo{person}{Karl~H Johansson}.}
  \bibinfo{year}{2015}\natexlab{}.
\newblock \showarticletitle{Heavy-duty vehicle platooning for sustainable
  freight transportation: A cooperative method to enhance safety and
  efficiency}.
\newblock \bibinfo{journal}{\emph{IEEE Control Systems Magazine}}
  \bibinfo{volume}{35}, \bibinfo{number}{6} (\bibinfo{year}{2015}),
  \bibinfo{pages}{34--56}.
\newblock


\bibitem[Dutta et~al\mbox{.}(2017)]%
        {dutta2017estimation}
\bibfield{author}{\bibinfo{person}{Raj~Gautam Dutta}, \bibinfo{person}{Xiaolong
  Guo}, \bibinfo{person}{Teng Zhang}, \bibinfo{person}{Kevin Kwiat},
  \bibinfo{person}{Charles Kamhoua}, \bibinfo{person}{Laurent Njilla}, {and}
  \bibinfo{person}{Yier Jin}.} \bibinfo{year}{2017}\natexlab{}.
\newblock \showarticletitle{Estimation of safe sensor measurements of
  autonomous system under attack}. In \bibinfo{booktitle}{\emph{Proceedings of
  the 54th Annual Design Automation Conference 2017}}. \bibinfo{pages}{1--6}.
\newblock


\bibitem[Han et~al\mbox{.}(2017)]%
        {han2017convoy}
\bibfield{author}{\bibinfo{person}{Jun Han}, \bibinfo{person}{Madhumitha
  Harishankar}, \bibinfo{person}{Xiao Wang}, \bibinfo{person}{Albert~Jin
  Chung}, {and} \bibinfo{person}{Patrick Tague}.}
  \bibinfo{year}{2017}\natexlab{}.
\newblock \showarticletitle{Convoy: Physical context verification for vehicle
  platoon admission}. In \bibinfo{booktitle}{\emph{Proceedings of the 18th
  International Workshop on Mobile Computing Systems and Applications}}.
  \bibinfo{pages}{73--78}.
\newblock


\bibitem[Jia et~al\mbox{.}(2015)]%
        {jia2015survey}
\bibfield{author}{\bibinfo{person}{Dongyao Jia}, \bibinfo{person}{Kejie Lu},
  \bibinfo{person}{Jianping Wang}, \bibinfo{person}{Xiang Zhang}, {and}
  \bibinfo{person}{Xuemin Shen}.} \bibinfo{year}{2015}\natexlab{}.
\newblock \showarticletitle{A survey on platoon-based vehicular cyber-physical
  systems}.
\newblock \bibinfo{journal}{\emph{IEEE communications surveys \& tutorials}}
  \bibinfo{volume}{18}, \bibinfo{number}{1} (\bibinfo{year}{2015}),
  \bibinfo{pages}{263--284}.
\newblock


\bibitem[Juuti et~al\mbox{.}(2017)]%
        {juuti2017stash}
\bibfield{author}{\bibinfo{person}{Mika Juuti}, \bibinfo{person}{Christian
  Vaas}, \bibinfo{person}{Ivo Sluganovic}, \bibinfo{person}{Hans Liljestrand},
  \bibinfo{person}{N Asokan}, {and} \bibinfo{person}{Ivan Martinovic}.}
  \bibinfo{year}{2017}\natexlab{}.
\newblock \showarticletitle{STASH: Securing transparent authentication schemes
  using prover-side proximity verification}. In \bibinfo{booktitle}{\emph{2017
  14th Annual IEEE International Conference on Sensing, Communication, and
  Networking (SECON)}}. IEEE, \bibinfo{pages}{1--9}.
\newblock


\bibitem[Kapoor et~al\mbox{.}(2018)]%
        {kapoor2018detecting}
\bibfield{author}{\bibinfo{person}{Prateek Kapoor}, \bibinfo{person}{Ankur
  Vora}, {and} \bibinfo{person}{Kyoung-Don Kang}.}
  \bibinfo{year}{2018}\natexlab{}.
\newblock \showarticletitle{Detecting and mitigating spoofing attack against an
  automotive radar}. In \bibinfo{booktitle}{\emph{2018 IEEE 88th Vehicular
  Technology Conference (VTC-Fall)}}. IEEE, \bibinfo{pages}{1--6}.
\newblock


\bibitem[Lioris et~al\mbox{.}(2017)]%
        {lioris2017platoons}
\bibfield{author}{\bibinfo{person}{Jennie Lioris}, \bibinfo{person}{Ramtin
  Pedarsani}, \bibinfo{person}{Fatma~Yildiz Tascikaraoglu}, {and}
  \bibinfo{person}{Pravin Varaiya}.} \bibinfo{year}{2017}\natexlab{}.
\newblock \showarticletitle{Platoons of connected vehicles can double
  throughput in urban roads}.
\newblock \bibinfo{journal}{\emph{Transportation Research Part C: Emerging
  Technologies}}  \bibinfo{volume}{77} (\bibinfo{year}{2017}),
  \bibinfo{pages}{292--305}.
\newblock


\bibitem[Lyamin et~al\mbox{.}(2016)]%
        {lyamin2016study}
\bibfield{author}{\bibinfo{person}{Nikita Lyamin}, \bibinfo{person}{Qichen
  Deng}, {and} \bibinfo{person}{Alexey Vinel}.}
  \bibinfo{year}{2016}\natexlab{}.
\newblock \showarticletitle{Study of the platooning fuel efficiency under {ETSI
  ITS-G5} communications}. In \bibinfo{booktitle}{\emph{Proc. of IEEE 19th
  ITSC}}. \bibinfo{pages}{551--556}.
\newblock


\bibitem[Maiti et~al\mbox{.}(2017)]%
        {maiti2017conceptualization}
\bibfield{author}{\bibinfo{person}{Santa Maiti}, \bibinfo{person}{Stephan
  Winter}, {and} \bibinfo{person}{Lars Kulik}.}
  \bibinfo{year}{2017}\natexlab{}.
\newblock \showarticletitle{A conceptualization of vehicle platoons and platoon
  operations}.
\newblock \bibinfo{journal}{\emph{Transportation Research Part C: Emerging
  Technologies}}  \bibinfo{volume}{80} (\bibinfo{year}{2017}),
  \bibinfo{pages}{1--19}.
\newblock


\bibitem[Rajamani(2011)]%
        {rajamani2011vehicle}
\bibfield{author}{\bibinfo{person}{Rajesh Rajamani}.}
  \bibinfo{year}{2011}\natexlab{}.
\newblock \bibinfo{booktitle}{\emph{Vehicle dynamics and control}}.
\newblock \bibinfo{publisher}{Springer Science \& Business Media}.
\newblock


\bibitem[Segata et~al\mbox{.}(2022)]%
        {segata2022multi}
\bibfield{author}{\bibinfo{person}{Michele Segata}, \bibinfo{person}{Renato~Lo
  Cigno}, \bibinfo{person}{Tobias Hardes}, \bibinfo{person}{Julian Heinovski},
  \bibinfo{person}{Max Schettler}, \bibinfo{person}{Bastian Bloessl},
  \bibinfo{person}{Christoph Sommer}, {and} \bibinfo{person}{Falko Dressler}.}
  \bibinfo{year}{2022}\natexlab{}.
\newblock \showarticletitle{Multi-Technology Cooperative Driving: An Analysis
  Based on PLEXE}.
\newblock \bibinfo{journal}{\emph{IEEE Transactions on Mobile Computing}}
  (\bibinfo{year}{2022}).
\newblock


\bibitem[Segata et~al\mbox{.}(2014)]%
        {segata2014plexe}
\bibfield{author}{\bibinfo{person}{Michele Segata}, \bibinfo{person}{Stefan
  Joerer}, \bibinfo{person}{Bastian Bloessl}, \bibinfo{person}{Christoph
  Sommer}, \bibinfo{person}{Falko Dressler}, {and} \bibinfo{person}{Renate~Lo
  Cigno}.} \bibinfo{year}{2014}\natexlab{}.
\newblock \showarticletitle{Plexe: A platooning extension for Veins}. In
  \bibinfo{booktitle}{\emph{2014 IEEE Vehicular Networking Conference (VNC)}}.
  IEEE, \bibinfo{pages}{53--60}.
\newblock


\bibitem[Shoukry et~al\mbox{.}(2015)]%
        {shoukry2015pycra}
\bibfield{author}{\bibinfo{person}{Yasser Shoukry}, \bibinfo{person}{Paul
  Martin}, \bibinfo{person}{Yair Yona}, \bibinfo{person}{Suhas Diggavi}, {and}
  \bibinfo{person}{Mani Srivastava}.} \bibinfo{year}{2015}\natexlab{}.
\newblock \showarticletitle{Pycra: Physical challenge-response authentication
  for active sensors under spoofing attacks}. In
  \bibinfo{booktitle}{\emph{Proceedings of the 22nd ACM SIGSAC Conference on
  Computer and Communications Security}}. \bibinfo{pages}{1004--1015}.
\newblock


\bibitem[Singh et~al\mbox{.}(2022)]%
        {singhVrange2022}
\bibfield{author}{\bibinfo{person}{Mridula Singh}, \bibinfo{person}{Marc
  Röschlin}, \bibinfo{person}{Aanjhan Ranganathan}, {and}
  \bibinfo{person}{Srdjan Capkun}.} \bibinfo{year}{2022}\natexlab{}.
\newblock \showarticletitle{V-Range: Enabling Secure Ranging in 5G Wireless
  Networks}. In \bibinfo{booktitle}{\emph{Proc. of the NDSS Symposium, to
  appear}}.
\newblock


\bibitem[Soltanaghaei et~al\mbox{.}(2018)]%
        {soltanaghaei2018multipath}
\bibfield{author}{\bibinfo{person}{Elahe Soltanaghaei},
  \bibinfo{person}{Avinash Kalyanaraman}, {and} \bibinfo{person}{Kamin
  Whitehouse}.} \bibinfo{year}{2018}\natexlab{}.
\newblock \showarticletitle{Multipath triangulation: Decimeter-level wifi
  localization and orientation with a single unaided receiver}. In
  \bibinfo{booktitle}{\emph{Proceedings of the 16th annual international
  conference on mobile systems, applications, and services}}.
  \bibinfo{pages}{376--388}.
\newblock


\bibitem[Turri et~al\mbox{.}(2016)]%
        {turri2016cooperative}
\bibfield{author}{\bibinfo{person}{Valerio Turri}, \bibinfo{person}{Bart
  Besselink}, {and} \bibinfo{person}{Karl~H Johansson}.}
  \bibinfo{year}{2016}\natexlab{}.
\newblock \showarticletitle{Cooperative look-ahead control for fuel-efficient
  and safe heavy-duty vehicle platooning}.
\newblock \bibinfo{journal}{\emph{IEEE Transactions on Control Systems
  Technology}} \bibinfo{volume}{25}, \bibinfo{number}{1}
  (\bibinfo{year}{2016}), \bibinfo{pages}{12--28}.
\newblock


\bibitem[Vaas et~al\mbox{.}(2018)]%
        {vaas2018get}
\bibfield{author}{\bibinfo{person}{Christian Vaas}, \bibinfo{person}{Mika
  Juuti}, \bibinfo{person}{N Asokan}, {and} \bibinfo{person}{Ivan Martinovic}.}
  \bibinfo{year}{2018}\natexlab{}.
\newblock \showarticletitle{Get in line: Ongoing co-presence verification of a
  vehicle formation based on driving trajectories}. In
  \bibinfo{booktitle}{\emph{2018 IEEE European Symposium on Security and
  Privacy (EuroS\&P)}}. IEEE, \bibinfo{pages}{199--213}.
\newblock


\bibitem[Waldschmidt et~al\mbox{.}(2021)]%
        {waldschmidt2021automotive}
\bibfield{author}{\bibinfo{person}{Christian Waldschmidt},
  \bibinfo{person}{Juergen Hasch}, {and} \bibinfo{person}{Wolfgang Menzel}.}
  \bibinfo{year}{2021}\natexlab{}.
\newblock \showarticletitle{Automotive radar—From first efforts to future
  systems}.
\newblock \bibinfo{journal}{\emph{IEEE Journal of Microwaves}}
  \bibinfo{volume}{1}, \bibinfo{number}{1} (\bibinfo{year}{2021}),
  \bibinfo{pages}{135--148}.
\newblock


\bibitem[Wang et~al\mbox{.}(2018)]%
        {wang2018review}
\bibfield{author}{\bibinfo{person}{Ziran Wang}, \bibinfo{person}{Guoyuan Wu},
  {and} \bibinfo{person}{Matthew~J Barth}.} \bibinfo{year}{2018}\natexlab{}.
\newblock \showarticletitle{A review on cooperative adaptive cruise control
  (CACC) systems: Architectures, controls, and applications}. In
  \bibinfo{booktitle}{\emph{2018 21st International Conference on Intelligent
  Transportation Systems (ITSC)}}. IEEE, \bibinfo{pages}{2884--2891}.
\newblock


\bibitem[Xu et~al\mbox{.}(2022)]%
        {xu2021pof}
\bibfield{author}{\bibinfo{person}{Ziqi Xu}, \bibinfo{person}{Jingcheng Li},
  \bibinfo{person}{Yanjun Pan}, \bibinfo{person}{Loukas Lazos},
  \bibinfo{person}{Ming Li}, {and} \bibinfo{person}{Nirnimesh Ghose}.}
  \bibinfo{year}{2022}\natexlab{}.
\newblock \showarticletitle{PoF: Proof-of-Following for Vehicle Platoons}. In
  \bibinfo{booktitle}{\emph{Proc. of the NDSS Symposium, to appear}}.
\newblock


\bibitem[Yeong et~al\mbox{.}(2021)]%
        {yeong2021sensor}
\bibfield{author}{\bibinfo{person}{De~Jong Yeong}, \bibinfo{person}{Gustavo
  Velasco-Hernandez}, \bibinfo{person}{John Barry}, \bibinfo{person}{Joseph
  Walsh}, {et~al\mbox{.}}} \bibinfo{year}{2021}\natexlab{}.
\newblock \showarticletitle{Sensor and sensor fusion technology in autonomous
  vehicles: A review}.
\newblock \bibinfo{journal}{\emph{Sensors}} \bibinfo{volume}{21},
  \bibinfo{number}{6} (\bibinfo{year}{2021}), \bibinfo{pages}{2140}.
\newblock


\end{thebibliography}

\end{document}